\documentclass[10pt,a4paper]{article}

\usepackage[top=3cm, left=3cm, bottom=4cm, right=2.5cm]{geometry} 
\usepackage{graphicx}  
\usepackage{amsmath, amsthm, amssymb}
\usepackage{overpic}
\usepackage{harvard}
\usepackage{enumerate}
\usepackage[normalem]{ulem}   
\usepackage[font=small,labelfont=bf]{caption}         
\DeclareMathAlphabet{\mathpzc}{OT1}{pzc}{m}{it}
\usepackage{xcolor}
\usepackage[ansinew]{inputenc}
\usepackage[T1]{fontenc}

\theoremstyle{plain}

\newtheorem{theorem}{Theorem}[section]
\newtheorem{proposition}[theorem]{Proposition}
\newtheorem{lemma}[theorem]{Lemma}
\newtheorem{corollary}[theorem]{Corollary}
\theoremstyle{definition}

\newtheorem{definition}[theorem]{Definition}
\newtheorem{assumption}[theorem]{Assumption}
\newtheorem{remark}[theorem]{Remark}
\newtheorem{example}[theorem]{Example}

\newcommand{\al}{\alpha}
\newcommand{\be}{\beta}
\newcommand{\ga}{\gamma}

\newcommand{\ka}{\kappa}
\newcommand{\ep}{\epsilon}
\newcommand{\T}{\Theta}
\newcommand{\tT}{\tilde{\Theta}}
\newcommand{\Tud}{\Theta^\updownarrow}
\newcommand{\s}{\sigma}
\newcommand{\de}{\delta}
\newcommand{\De}{\Delta}
\def\cA{\mathcal{A}} 
\def\ctA{\tilde{\mathcal{A}}} 
\def\cB{\mathcal{B}} 
\def\cC{\mathcal{C}}
\def\cF{\mathcal{F}} 
\def\cH{\mathcal{H}} 
\def\cP{\mathcal{P}} 
\def\N{\mathbb{N}}   
\def\R{\mathbb{R}}   
\def\E{\mathbb{E}}   
\def\tD{E}   

\title{Optimal trade execution and price manipulation in order books with time-varying liquidity\footnote{We would like to thank Peter Bank for valuable suggestions.}}
\author{Antje Fruth\thanks{Technische Universität Berlin, Germany, fruth@math.tu-berlin.de} \and 
Torsten Schöneborn\thanks{Deutsche Bank AG, London, UK, schoeneborn@math.tu-berlin.de} \and
Mikhail Urusov\thanks{Ulm University, Germany, mikhail.urusov@uni-ulm.de}}
\date{\today}

\begin{document}
\maketitle
\begin{abstract}
In financial markets, liquidity is not constant over time but exhibits strong seasonal patterns. In this article we consider a limit order book model that allows for time-dependent, deterministic depth and resilience of the book and determine optimal portfolio liquidation strategies. In a first model variant, we propose a trading dependent spread that increases when market orders are matched against the order book. In this model no price manipulation occurs and the optimal strategy is of the wait region - buy region type often encountered in singular control problems. In a second model, we assume that there is no spread in the order book. Under this assumption we find that  price manipulation can occur, depending on the model parameters. Even in the absence of classical price manipulation there may be transaction triggered price manipulation. In specific cases, we can state the optimal strategy in closed form.
\end{abstract}


\section{Introduction}\label{SecIntroduction}

Empirical investigations have demonstrated that liquidity varies over time. In particular deterministic time-of-day and day-of-week liquidity patterns have been found in most markets, see, e.g., \citeasnoun{Chordia2001}, \citeasnoun{Kempf2008} and \citeasnoun{LorenzOsterrieder}. In spite of these findings the academic literature on optimal trade execution usually assumes constant liquidity during the trading time horizon. 
In this paper we relax this assumption and analyze the effects of deterministically\footnote{Not all changes in liquidity are deterministic; an additional stochastic component has been investigated empirically by, e.g., \citeasnoun{EsserMonch} and \citeasnoun{Steinmann2005}. See, e.g., \citeasnoun{Diss} for an analysis of the implications of such stochastic liquidity on optimal trade execution.} varying liquidity on optimal trade execution for a risk-neutral investor. We characterize optimal strategies in terms of a trade region and a wait region and find that optimal trading strategies depend on the expected pattern of time-dependent liquidity. In the case of extreme changes in liquidity, it can even be optimal to entirely refrain from trading in periods of low liquidity. Incorporating such patterns in trade execution models can hence lower transaction costs.

Time-dependent liquidity can potentially lead to price manipulation. In periods of low liquidity, a trader could buy the asset and push market prices up significantly; in a subsequent period of higher liquidity, he might be able to unwind this long position without depressing market prices to their original level, leaving the trader with a profit after such a round trip trade. In reality such round trip trades are often not profitable due to the bid-ask spread: once the trader starts buying the asset in large quantities, the spread widens, resulting in a large cost for the trader when unwinding the position. We propose a model with trading-dependent spread and demonstrate that price manipulation does not exist in this model in spite of time-dependent liquidity. In a similar model with fixed zero spread we find that price manipulation or transaction-triggered price manipulation (a term recently coined by \citeasnoun{ASS} and \citeasnoun{GatheralSchiedSlynko1}) can be a consequence of time-dependent liquidity. Phenomena of such type, i.e. existence of ``illusory arbitrages'', which disappear when bid-ask spread is taken into account, are also observed in different modelling approaches (see e.g. Section 5.1 in \citeasnoun{MadanSchoutens}).

Our liquidity model is based on the limit order book market model of \citeasnoun{OW}, which models both depth and resilience of the order book explicitly. The instantaneously available liquidity in the order book is described by the depth. Market orders issued by the large investor are matched with this liquidity, which increases the spread. Over time, incoming limit orders replenish the order book and reduce the spread; the speed of this process is determined by the resilience. In our model both depth and resilience can be independently time dependent. We show that there is a time dependent optimal ratio of remaining order size to bid-ask spread: If the actual ratio is larger than the optimal ratio, then the trader is in the ``trade region'' and it is optimal to reduce the ratio by executing a part of the total order. If the actual ratio is smaller than the optimal ratio, then the trader is in the ``wait region'' and it is optimal to wait for the spread to be reduced by future incoming limit orders before continuing to trade. 

Building on empirical investigations of the market impact of large transactions, a number of theoretical models of illiquid markets have emerged. One part of these market microstructure models focuses on the underlying mechanisms for illiquidity effects, e.g., \citeasnoun{Kyle} and \citeasnoun{EasleyOHara}. We follow a second line that takes the liquidity effects as given and derives optimal trading strategies within such a stylized model market. Two broad types of market models have been proposed for this purpose. First, several models assume an instantaneous price impact, e.g., \citeasnoun{BertsimasLo}, \citeasnoun{AlmgrenChriss2001} and \citeasnoun{Almgren2003}. The instantaneous price impact typically combines depth and resilience of the market into one stylized quantity. Time-dependent liquidity in this setting then leads to executing the constant liquidity strategy in volume time or liquidity time, and no qualitatively new features occur. In a second group of models resilience is finite and depth and resilience are separately modelled, e.g., \citeasnoun{Bouchaudetal}, \citeasnoun{OW}, \citeasnoun{AFS1} and \citeasnoun{PredoiuShaikhetShreve}. Our model falls into this last group. Allowing for independently time-dependent depth and resilience leads to higher technical complexity, but allows us to capture a wider range of real world phenomena.

The remainder of this paper is structured as follows. In the next section, we introduce the market model and formulate an optimization problem. In Section~\ref{SecMarketManipulation}, we show that this model is free of price manipulation, which allows us to simplify the model setup and the optimization problem in Section~\ref{sec:red_opt_prob}. Before we state our main results on existence, uniqueness and characterization of the optimal trading strategy in Sections~\ref{Subsec: WR-BR deterministic} to~\ref{Subsec: WR-BR deterministic cts}, we first provide some elementary properties, like the dimension reduction of our control problem, in Section~\ref{Subsec: Preparations}. Section~\ref{Subsec: WR-BR deterministic} discusses the case where trading is constrained to discrete time and Section~\ref{Subsec: WR-BR deterministic cts} contains the continuous time case. In Section~\ref{SecZeroSpreadPriceManipulation} we investigate under which conditions price manipulation occurs in a zero spread model. In some special cases, we can calculate optimal strategies in closed form for our main model as well as for the zero spread model of Section~\ref{SecZeroSpreadPriceManipulation}; we provide some examples in Section~\ref{Subsec: Euler Lagrange}.
Section~\ref{sec:conclusion} concludes.
\section{Model description}\label{SecModel}

In order to attack the problem of optimal trade execution under time-varying liquidity, we first need to specify a price impact model in Section \ref{Subsec: LOB model}. Our model is based on the work of \citeasnoun{OW}, but allows for time-varying order book depth and resilience. Furthermore we explicitly model both sides of the limit order book and thus can allow for strategies that buy and sell at different points in time. After having introduced the limit order book model, we specify the trader's objectives in Section \ref{Subsec: Trader objectives}.


\subsection{Limit order book and price impact}\label{Subsec: LOB model}
Trading at most public exchanges is executed through a limit order book, which is a collection of the limit orders of all market participants in an electronic market. Each limit order has the number of shares, that the market participant wants to trade, and a price per share attached to it. The price represents a minimal price in case of a sell and a maximal price in case of a buy order. Compared to a limit order, a market order does not have an attached price per share, but instead is executed immediately against the best limit orders waiting in the book. Thus, there is a tradeoff between price saving and immediacy when using limit and market orders. We refer the reader to \citeasnoun{ContStoikovTalreja} for a more comprehensive introduction to limit order books. 

In this paper we consider a one-asset model that derives its price dynamics from a limit order book that is exposed to repeated market orders of a large investor (sometimes referred to as the trader). The goal of the investor is to use market orders\footnote{On this macroscopic time scale, the restriction to market orders is not severe. A subsequent consideration of small time windows including limit order trading is common practice in banks. See \citeasnoun{Naujokat} for a discussion of a large investor execution problem where both market and limit orders are allowed.} in order to purchase a large amount~$x$ of shares within a certain time period~$[0,T]$, where~$T$ typically ranges from a few hours up to a few trading days. Without loss of generality we assume that the investor needs to purchase the asset (the sell case is symmetrical) and hence first describe how buy market orders interact with the ask side of the order book (i.e., with the sell limit orders contained in the limit order book). Subsequently we turn to the impact of buy market orders on the bid side and of sell market orders on both sides of the limit order book. 

Suppose first that the trader is not active. We assume that the corresponding unaffected best ask price~$A^u$ (i.e. the lowest ask price in the limit order book) is a c\`adl\`ag martingale on a given filtered probability space~$(\Omega,\cF,(\cF_t)_{t \in [0,T]},\mathbb{P})$ satisfying the usual conditions. This unaffected price process is capturing all external price changes including those due to news as well as due to trading of noise traders and informed traders. Our model includes in particular the case of the Bachelier model~$A_t^u=A^u_0+\sigma W^A_t$ with a~$(\cF_t)$-Brownian motion~$W^A$, as considered in~\citeasnoun{OW}. It also includes the driftless geometric Brownian motion~$A_t^u=A^u_0 \exp(\s W^A_t - \frac{1}{2}\s^2 t)$, which avoids the counterintuitive negative prices of the Bachelier model. Moreover, we can allow for jumps in the dynamics of~$A^u$.

\begin{figure}[tbp]
 \centering
 \includegraphics[width=0.7\linewidth]{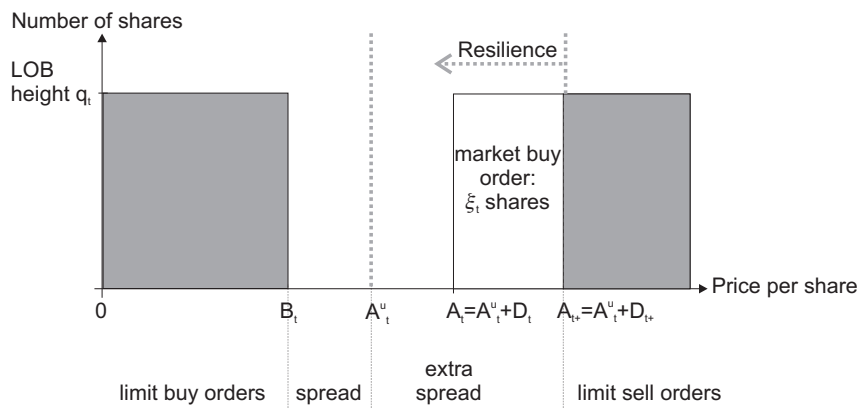} 
 \caption{Snapshot of the block-shaped order book model at time~$t$.}
 \label{LOB figure}
\end{figure}

We now describe the shape of the limit order book, i.e. the pattern of ask prices in the order book. We follow \citeasnoun{OW} and assume a block-shaped order book: The number of shares offered at prices in the interval~ $[A_t^u,A_t^u+\De A]$ is given by~${q_t}\cdot \De A$ with ${q_t}>0$ being the order book height (see Figure~\ref{LOB figure} for a graphical illustration). 
\citeasnoun{AFS1} and \citeasnoun{PredoiuShaikhetShreve} consider order books which are not block shaped and conclude that the optimal execution strategy of the investor is robust with respect to the order book shape. In our model, we allow the order book depth~$q_t$ to be time dependent. As mentioned above, various empirical studies have demonstrated the time-varying features of liquidity, including order book depth. In theoretical models however, liquidity is still usually assumed to be constant in time. To our knowledge first attempts to non-constant liquidity in portfolio liquidation problems has only been considered so far in extensions of the \citeasnoun{AlmgrenChriss2001} model such as \citeasnoun{KimBoyd} and \citeasnoun{AlmgrenDynamic}. In this modelling framework, price impact is purely temporary and several of the aspects of this paper do not surface. 

Let us now turn to the interaction of the investor's trading with the order book. At time $t$, the best ask $A_t$ might differ from the unaffected best ask $A_t^u$ due to previous trades of the investor. Define $D_t:=A_t-A_t^u$ as the \emph{price impact} or \emph{extra spread} caused by the past actions of the trader. Suppose that the trader places a buy market order of~$\xi_t>0$ shares. This market order consumes all the shares offered at prices between the ask price~$A_t$ just prior to order execution and $A_{t+}$ immediately after order execution. $A_{t+}$ is given by $\left(A_{t+}-A_t\right) \cdot {q_t}=\xi_t$ and we obtain $$D_{t+}=D_t + \xi_t/q_t.$$ See Figure~\ref{LOB figure} for a graphical illustration.

It is a well established empirical fact that the price impact $D$ exhibits  resilience over time. We assume that the immediate impact~$\xi_t/q_t$ can be split into a temporary impact component $K_t \xi_t$ which decays to zero and a permanent impact component $\ga \xi_t$ with
$$\ga+K_t=q_t^{-1}.$$
We assume that the temporary impact decays exponentially with a fixed time-dependent, deterministic recovery rate $\rho_t>0$. The price impact at time~$s \geq t$ of a buy market order~$\xi_t>0$ placed at time~$t$ is assumed to be
$$\ga \xi_t+K_t e^{-\int_t^{s}\rho_u\,du}\xi_t.$$
Notice that this temporary impact model is different to the one which is used, e.g.,~in \citeasnoun{AlmgrenChriss2001} and \citeasnoun{Almgren2003}. It slowly decays to zero instead of vanishing immediately and thus prices depend on previous trades. 
\citeasnoun{OW} limit their analysis to a constant decay rate $\rho_t \equiv \rho$, but suggest the extension to time dependent $\rho_t$. \citeasnoun{Weiss} considers exponential resilience and shows that the results of \citeasnoun{AFS1} and in particular \citeasnoun{OW} can be adapted when the recovery rate depends on the extra spread~$D$ caused by the large investor. \citeasnoun{Gatheral} considers more general deterministic decay functions than the exponential one in a model with a potentially non-linear price impact and discusses which combinations of decay function and price impact yield 'no arbitrage', i.e. non-negative expected costs of a round trip. \citeasnoun{ASS} study the optimal execution problem for more general deterministic decay functions than the exponential one in a model with constant order book height. For the calibration of resilience see \citeasnoun{Large} and for a discussion of a stochastic recovery rate~$\rho$ we refer to \citeasnoun{Diss}. 

Let us now discuss the impact of market buy orders on the bid side of the limit order book. According to the mechanics of the limit order book, a single market buy order~$\xi_t$ directly influences the best ask~$A_{t+}$, but does not influence the best bid price~$B_{t+}=B_t$ immediately. The best ask~$A_{t+}$ recovers over time (in the absence of any other trading from the investor) on average to~$A_t + \ga \xi_t$. In reality market orders only lead to a temporary widening of the spread. In order to close the spread,~$B_t$ needs to move up by~$\ga \xi_t$ over time and converge to~$B_t + \ga \xi_t$, i.e. the buy market order~$\xi_t$ influences the future evolution of $B$. We assume that~$B$ converges to this new level exponentially with the same rate~$\rho_t$. The price impact on the best bid~$B_s$ at time~$s \geq t$ of a buy market order~$\xi_t>0$ placed at time~$t$ is hence
$$\ga \left(1-e^{-\int_t^{s}\rho_u\,du} \right)\xi_t.$$
We assume that the impact of sell market orders is symmetrical to that of buy market orders. It should be noted that our model deviates from the existing literature by explicitly modelling both sides of the order book with a trading dependent spread. For example \citeasnoun{OW} only model one side of the order book and restrict trading to this side of the book. 
\citeasnoun{ASS}, \citeasnoun{GatheralSchiedSlynko1} and \citeasnoun{GatheralSchiedSlynko2} on the other hand allow for trading on both sides of the order book, but assume that there is no spread, i.e. they assume $A^u=B^u$ for unaffected best ask and best bid prices, and that the best bid moves up instantaneously when a market buy order is matched with the ask side of the book. They find that under this assumption the model parameters (for example the decay kernel) need to fulfill certain conditions, otherwise price manipulation arises. We will revisit this topic in Sections \ref{SecMarketManipulation} and \ref{SecZeroSpreadPriceManipulation}.

We can now summarize the dynamics of the best ask $A_t$ and best bid $B_t$ for general trading strategies in continuous time. Let~$\T$ and $\tT$ be increasing processes that describe the number of shares which the investor bought respectively sold from time $0$ until time~$t$. We then have
\begin{eqnarray*}
	A_t & = &  A_t^u+D_t,\\
	B_t & = &  B_t^u - \tD_t,
\end{eqnarray*}
where
\begin{align}\label{D integral form 2}
	D_t & =  D_0 e^{-\int_0^t \rho_s ds}+\int_{[0,t)} \left( \ga+K_s e^{-\int_s^t \rho_u du}\right) d\T_s 
	- \int_{[0,t)} \ga \left(1-e^{-\int_s^t \rho_u du}\right) d\tT_s 	, \quad t \in [0,T+],\\
	\tD_t & =  \tD_0 e^{-\int_0^t \rho_s ds}+\int_{[0,t)} \left( \ga+K_s e^{-\int_s^t \rho_u du}\right) d\tT_s 
	- \int_{[0,t)} \ga \left(1-e^{-\int_s^t \rho_u du}\right) d\T_s 	, \quad t \in [0,T+],
\end{align}
with some given nonnegative initial price impacts $D_0\ge0$ and $\tD_0\ge0$.

\begin{assumption}[Basic assumptions on $\T$, $\tT$, $A^u$, $B^u$, $K$, and~$\rho$]
\label{basic assumption}\mbox{}\\
Throughout this paper, we assume the following.

\vspace{-4mm}\noindent
\begin{itemize}
\item The set of \emph{admissible strategies} is given as
\begin{align*}
	\ctA_0&:=\big\{(\T, \tT)\colon\Omega \times [0,T+] \rightarrow [0,\infty)^2 \,|\,
	\T\text{ and }\tT\text{ are }(\cF_t)\text{-adapted nondecreasing}\\
	 &\hspace{9mm}\text{bounded c\`agl\`ad processes with } (\T_0,\tT_0)=(0,0) \big\}.
\end{align*}
Note that~$(\T,\tT)$ may have jumps. In particular, trading in rates and impulse trades are allowed. 
\item The unaffected best ask price process $A^u$ is a c\`adl\`ag $\cH^1$-martingale with a deterministic starting point~$A^u_0$, i.e.
$$
\E\sqrt{[A^u,A^u]_T} < \infty,\quad\text{or, equivalently,}\quad\E\sup_{t \in [0,T]}{|A^u_t|} < \infty.
$$
The same condition holds for the unaffected best bid price~$B^u$. Furthermore, $B^u_t \leq A^u_t$ for all~$t \in [0,T]$.
\item The price impact coefficient $K\colon[0,T] \rightarrow (0,\infty)$ is a deterministic strictly positive bounded Borel function.
\item The resilience speed $\rho\colon[0,T]\rightarrow (0,\infty)$ is a deterministic strictly positive Lebesgue integrable function.
\end{itemize} 
\end{assumption}

\begin{remark}\mbox{}

\vspace{-4mm}\noindent
\begin{enumerate}[i)]
\item The purchasing component $\T$ of a strategy from $\ctA_0$ consists of a left-continuous nondecreasing process $(\T_t)_{t \in [0,T]}$ and an additional random variable $\T_{T+}$ with $\De \T_T:=\T_{T+}-\T_{T} \geq 0$ being the last purchase of the strategy. Similarly, for~$t \in [0,T]$, we use the notation $\De \T_t:=\T_{t+}-\T_t$. The same conventions apply for the selling component~$\tT$.
	\item The processes $D$ and $E$ depend on $(\T,\tT)$, although this is not explicitly marked in their notation.
	\item As it is often done in the literature on optimal portfolio execution,~$\T$, $\tT$, $D$ and $E$ are assumed to be c\`agl\`ad processes. In~(\ref{D integral form 2}), the possibility~$t=T+$ is by convention understood as 
	$$D_{T+}=D_0 e^{-\int_0^T \rho_s ds}+\int_{[0,T]} \left( \ga+K_s e^{-\int_s^T \rho_u du}\right) d\T_s 
		- \int_{[0,T]} \ga e^{-\int_s^T \rho_u du} d\tT_s .$$
	A similar convention applies to all other formulas of such type. Furthermore, the integrals of the form
	$$\int_{[0,t)} K_s d \T_s \hspace{.2cm} \text{or} \hspace{.2cm} \int_{[0,t]} K_s d \T_s,$$
	are understood as pathwise Lebesgue-Stieltjes integrals, i.e. Lebesgue integrals with respect to the measure with the distribution function~$s \mapsto \T_{s+}$.
	\item In the sequel, we need to apply stochastic analysis (e.g. integration by parts or Ito's formula) to c\`agl\`ad processes of finite variation and/or standard semimartingales. This will always be done as follows: if~$U$ is a c\`agl\`ad process of finite variation, we first consider the process~$U^+$ defined by~$U^+_t:=U_{t+}$ and then apply standard formulas from stochastic analysis to it. An example (which will be often used in proofs) is provided in Appendix \ref{AppendixStochAna}.
\end{enumerate}
\end{remark}


\subsection{Optimization problem}\label{Subsec: Trader objectives}
Let us go ahead by describing the cost minimization problem of the trader. When placing a single buy market order of size~$\xi_t \ge 0$ at time~$t$, he purchases at prices~$A^u_t+d$, with~$d$ ranging from~$D_t$ to~$D_{t+}$, see Figure~\ref{LOB figure}. Due to the block-shaped limit order book, the total costs of the buy market order amount to
$$
\left(A^u_t+D_t\right) \xi_t+\frac{D_{t+}-{D_t}}{2} \xi_t=\left(A^u_t+D_t\right) \xi_t+\frac{\xi^2_t}{2 q_t}
=\xi_t\left(A_t+\frac{\xi_t}{2q_t}\right).
$$
Thus, the total costs of the buy market order are the number of shares $\xi_t$ times the average price per share~$(A_t+\frac{\xi_t}{2 q_t})$.
More generally, the total costs of a strategy $(\T,\tT)\in\ctA_0$ are given by the formula
$$
\cC(\T,\tT):=\int_{[0,T]} \left( A_t+\frac{\De \T_t}{2 q_t} \right) d\T_t 
		-		\int_{[0,T]} \left( B_t -\frac{\De \tT_t}{2 q_t} \right) d\tT_t.
$$

We now collect all admissible strategies that build up a position of $x \in [0,\infty)$ shares until time~$T$ in the set
\begin{equation*}
	\ctA_0(x):=\left \{ (\T,\tT) \in \ctA_0 \,|\, \T_{T+}-\tT_{T+}=x \text{ a.s.} \right\}.
\end{equation*}
Our aim is to minimize the expected execution costs
\begin{equation}\label{problem with sells}
		\inf_{(\T,\tT) \in \ctA_0(x)}\E\,\cC(\T,\tT).
\end{equation}
We hence consider the large investor to be risk-neutral and explicitly allow for his optimal strategy to consist of both buy and sell orders. In the next section, we will see that in our model it is never optimal to submit sell orders when the overall goal is the purchase of $x>0$ shares.

Let us finally note that problem~\eqref{problem with sells} with $x\in(-\infty,0]$
is the problem of maximizing the expected proceeds from liquidation of $|x|$ shares
and, due to symmetry in modelling ask and bid sides, can be considered similarly
to problem~\eqref{problem with sells} with $x\in[0,\infty)$.

\section{Market manipulation}\label{SecMarketManipulation}
Market manipulation has been a concern for price impact models for some time.
We now define the counterparts in our model of the notions of \emph{price manipulation}
in the sense of \citeasnoun{Huberman2004} and of \emph{transaction-triggered price manipulation}
in the sense of \citeasnoun{ASS} and \citeasnoun{GatheralSchiedSlynko1}.
Note that in defining these notions in our model we explicitly account
for the possibility of $D_0$ and $\tD_0$ being nonzero.

\begin{definition}
A \emph{round trip} is a strategy from~$\ctA_0(0)$.
A \emph{price manipulation strategy} is a round trip $(\T,\tT)\in\ctA_0(0)$ with strictly negative expected execution costs $\E\,\cC(\T,\tT)<0$.
A market impact model (represented by $A^u$, $B^u$, $K$, and~$\rho$)
admits \emph{price manipulation} if there exist $D_0\ge0$, $\tD_0\ge0$
and $(\T,\tT)\in\ctA_0(0)$ with $\E\,\cC(\T,\tT)<0$.
\end{definition}

\begin{definition}
\label{def:ttpm1}
A market impact model (represented by $A^u$, $B^u$, $K$, and~$\rho$)
admits \emph{transaction-triggered price manipulation}
if the expected execution costs of a buy (or~sell) program
can be decreased by intermediate sell (resp.~buy) trades.
More precisely, this means that there exist $x\in[0,\infty)$, $D_0\ge0$, $\tD_0\ge0$
and $(\T^0,\tT^0)\in\ctA_0(x)$ with
\begin{equation}
\label{eq:ttpm1}
\E\,\cC(\T^0,\tT^0)<\inf\{\E\,\cC(\T,0)\,|\,(\T,0)\in\ctA_0(x)\}
\end{equation}
or there exist $x\in(-\infty,0]$, $D_0\ge0$, $\tD_0\ge0$
and $(\T^0,\tT^0)\in\ctA_0(x)$ with
\begin{equation}
\label{eq:ttpm2}
\E\,\cC(\T^0,\tT^0)<\inf\{\E\,\cC(0,\tT)\,|\,(0,\tT)\in\ctA_0(x)\}.
\end{equation}
\end{definition}

Clearly, if a model admits price manipulation, then it admits transaction-triggered price manipulation.
But transaction-triggered price manipulation can be present even if price manipulation does not exist in a model. This situation has been demonstrated in limit order book models with zero bid-ask spread by \citeasnoun{DissTorsten} (Chapter~9) in a multi-agent setting and by \citeasnoun{ASS} in a setting with non-exponential decay of price impact. In this section, we will show that the limit order book model introduced in Section~\ref{SecModel} is free from both classical and transaction-triggered price manipulation. In Section~\ref{SecZeroSpreadPriceManipulation} we will revisit this topic for a different (but related) limit order book model.

Before attacking the main question of price manipulation in Proposition~\ref{only buy orders}, we consider the expected execution costs of a pure purchasing strategy and verify in Proposition~\ref{only temp impact} that the costs resulting from changes in the unaffected best ask price are zero and that the costs due to permanent impact are the same for all strategies.

\begin{proposition}[Only temporary impact has to be considered]
\label{only temp impact}\mbox{}\\
Let~$(\T,\tT) \in \ctA_0(x)$ with $x\in[0,\infty)$ and~$\tT \equiv 0$. Then
\begin{eqnarray}
\label{EqnReduction1}
\mathbb{E}\left[ \int_{[0,T]} \left(A_t+\frac{\De \T_t}{2 q_t} \right) d\T_t \right] =	
A^u_0 x+\frac{\ga}{2}x^2+\mathbb{E}	\left[ \int_{[0,T]} \left(	D^{\ga=0}_t+\frac{K_t }{2} \De \T_t \right) d\T_t \right]
\end{eqnarray}
with
\begin{equation}
\label{eq:D_ga0}
D^{\ga=0}_t := D_0 e^{-\int_0^t \rho_s ds}+\int_{[0,t)} K_s e^{-\int_s^t \rho_u du} d\T_s , \hspace{.2cm} t \in [0,T+].
\end{equation}
\end{proposition}

\begin{proof}
	We start by looking at the expected costs caused by the unaffected best ask price martingale. Using~(\ref{int by parts 1}) with~$U:=\T, \, Z:=A^u$, 
	the facts that~$\T$ is bounded and that~$A^u$ is an~$\cH^1$-martingale yield
	\begin{equation}\label{A reduction}
		\mathbb{E}\left[\int_{[0,T]} A^u_t d\T_t \right]=\mathbb{E}\left[ A^u_{T} \T_{T+}-A^u_0 \T_0 \right]=A^u_0 x.
	\end{equation}
	Let us now turn to the simplification of our optimization problem due to permanent impact. To this end, we differentiate between the temporary price 
	impact~$D^{\ga=0}_t$ and the total price impact~$D_t=D^{\ga=0}_t+\ga \T_t$ that we get by adding the permanent impact. Notice that~$\tT \equiv 0$. We can then write
	\begin{eqnarray*}
	  && \mathbb{E}\left[ \int_{[0,T]} \left(A^u_t+D_t+\frac{\De \T_t}{2 q_t} \right) d\T_t \right]\\
	  &=& A^u_0 x+\mathbb{E}\left[ \int_{[0,T]} \left(D^{\ga=0}_t+\ga \T_t+\frac{\ga+K_t}{2} \De \T_t  \right) d\T_t \right]\\
	  &=& A^u_0 x+\mathbb{E}\left[ \int_{[0,T]} \left(D^{\ga=0}_t+\frac{K_t}{2} \De \T_t  \right) d\T_t \right] 
	           + \ga \mathbb{E}\left[ \int_{[0,T]} \left( \T_t+\frac{\De \T_t}{2} \right) d\T_t \right].
	\end{eqnarray*}
	The assertion follows, since integration by parts for c\`agl\`ad processes (see~(\ref{int by parts 2}) with~$U=V:=\T$) and~$\T_0=0$,~$\T_{T+}=x$ yield
	\begin{equation}\label{x square in by parts}
		\int_{[0,T]} \left(\T_t+\frac{\De \T_t}{2} \right) d\T_t=\frac{\T^2_{T+}-\T^2_0}{2}=\frac{x^2}{2}.
	\end{equation}
\end{proof}

We can now proceed to prove that our model is free of price manipulation and transaction-triggered price manipulation.

\begin{proposition}[Absence of transaction-triggered price manipulation]
\label{only buy orders}\mbox{}\\
In the model of Section~\ref{SecModel}, there is no transaction-triggered price manipulation.
In particular, there is no price manipulation.
\end{proposition}

\begin{proof}
	Consider $x\in[0,\infty)$ and $(\T,\tT) \in \ctA_0(x)$. Making use of
	$$B_t=B_t^u-\tD_t \leq A_t^u-\tD_t \leq A_t^u + \ga \left(\T_t-\tT_t \right)$$
	yields
	\begin{eqnarray*}
		&&\mathbb{E} \left[ \int_{[0,T]} \left( A_t+\frac{\De \T_t}{2 q_t} \right) d\T_t \right]	- \mathbb{E} \left[ 
		\int_{[0,T]} \left( B_t -\frac{\De \tT_t}{2 q_t} \right) d \tT_t \right]\\
		&\geq& \mathbb{E} \left[ \int_{[0,T]} \left( A^u_t+\ga \T_t + D^{\ga=0}_t -\ga \tT_t + \frac{\ga}{2}\De \T_t+\frac{K_t}{2}\De \T_t \right) d\T_t \right]\\
		&&- \mathbb{E} \left[ \int_{[0,T]} \left(A^u_t+\ga \T_t-\ga \tT_t-\frac{\ga}{2}\De \tT_t-\frac{K_t}{2}\De \tT_t \right) 
		d \tT_t \right]\\
		&\geq& \mathbb{E} \left[ \int_{[0,T]} A_t^u d(\T_t-\tT_t) \right ]\\
		&&+ \ga \mathbb{E} \left[ \int_{[0,T]} \left(\T_t - \tT_t + \frac{\De \T_t}{2} \right) d\T_t+\int_{[0,T]}\left( \tT_t-\T_t+\frac{\De \tT_t}{2} 
		\right) d\tT_t \right]\\
		&&+ \mathbb{E} \left[\int_{[0,T]} \left( D^{\ga=0}_t+\frac{K_t}{2} \De \T_t \right) d\T_t \right].
	\end{eqnarray*}
Analogously to~\eqref{A reduction}, the first of these terms equals $A^u_0 x$ since $\T,\tT$ are bounded and $A^u$ is an $\cH^1$-martingale.
	For the second one, we do integration by parts (use~(\ref{int by parts 2}) three times) to deduce
	\begin{eqnarray*}
		&&\int_{[0,T]} \left(2 \T_t+\De \T_t \right) d\T_t+\int_{[0,T]}\left( 2 \tT_t+\De \tT_t \right) d\tT_t
		-2\int_{[0,T]}\tT_t d\T_t - 2\int_{[0,T]}\T_t d\tT_t\\
		&=& \T_{T+}^2+\tT_{T+}^2-2  \T_{T+} \tT_{T+} -2\int_{[0,T]}\tT_t d\T_t +2 \int_{[0,T]}\tT_{t+} d\T_t \geq \left(\T_{T+}- \tT_{T+} \right)^2=x^2.
	\end{eqnarray*}
	That is we have shown
	\begin{eqnarray*}
		&&\mathbb{E} \left[ \int_{[0,T]} \left( A_t+\frac{\De \T_t}{2 q_t} \right) d\T_t -
		\int_{[0,T]} \left( B_t -\frac{\De \tT_t}{2 q_t} \right) d \tT_t \right]\\
		&\geq& A^u_0 x+\frac{\ga}{2} x^2 + \mathbb{E} \left[\int_{[0,T]} \left( D^{\ga=0}_t+\frac{K_t}{2} \De \T_t \right) d\T_t \right]
	\end{eqnarray*}
	and thanks to Proposition~\ref{only temp impact} the right-hand side is larger or equal to the expected execution costs of the strategy~$(\check{\T},0) \in  
	\ctA_0(x)$ with
	$$\check{\T}_t:=\left\{ \begin{array}{cl} \T_t & \text{if } \T_t \leq x\\ x	&	\text{otherwise} \end{array}\right  \}.$$
	Hence, the expected execution costs of a buy program~$(\T,\tT) \in \ctA_0(x)$ containing a selling component is always greater or equal to the costs of the modified 
	strategy~$(\check{\T},0) \in  \ctA_0(x)$ without a selling component.
	Thus, \eqref{eq:ttpm1} does not occur. By a similar reasoning, \eqref{eq:ttpm2} does not occur as well.
\end{proof}

The central economic insight captured in the previous proposition is that price manipulation strategies can be severely penalized by a widening spread. This idea can easily be applied to different variations of our model, for example to non-exponential decay kernels as in \citeasnoun{GatheralSchiedSlynko1}.

\section{Reduction of the optimization problem}\label{sec:red_opt_prob}
Due to Propositions~\ref{only temp impact} and~\ref{only buy orders}, we can significantly simplify the optimization problem~\eqref{problem with sells}. Let us fix $x\in[0,\infty)$.
Then it is enough to minimize the expectation in the right-hand side of~\eqref{EqnReduction1}
over the pure purchasing strategies that build up the position of $x$ shares until time~$T$.
That is to say, the problem in general reduces to that with $A^u\equiv0$, $\ga=0$, $\tT\equiv0$.
Moreover, due to~\eqref{EqnReduction1},~\eqref{eq:D_ga0} and the fact that $K$ and $\rho$
are deterministic functions, it is enough to minimize over deterministic purchasing strategies.
We are going to formulate the simplified optimization problem, where we now consider a general initial time $t\in[0,T]$
because we will use dynamic programming afterwards.

Let us define the following simplified control sets only containing deterministic purchasing strategies:
\begin{align*}
\cA_t&:=\big\{\T\colon[t,T+]\rightarrow[0,\infty)\,|\,\T
\text{ is a deterministic}\\
&\hspace{9mm}\text{nondecreasing c\`agl\`ad function with }\T_t=0\big\},\\
\cA_t(x)&:=\left\{\T\in\cA_t\,|\,\T_{T+}=x\right\}.
\end{align*}
As above, a strategy from $\cA_t$ consists of a left-continuous nondecreasing function
$(\T_s)_{s \in [t,T]}$ and an additional value $\T_{T+}\in[0,\infty)$ with $\De\T_T:=\T_{T+}-\T_{T}\geq0$
being the last purchase of the strategy.
For any fixed $t\in[0,T]$ and $\de\in[0,\infty)$, we define the \emph{cost function}
$J(t,\de,\cdot)\colon\cA_t\rightarrow[0,\infty)$ as
\begin{equation}
\label{J}
J(\T):=J(t,\de,\T):= \int_{[t,T]} \left(D_s+\frac{K_s}{2} \De \T_s \right) d\T_s,
\end{equation}
where
\begin{equation}
\label{D explicit}
D_s:=\de e^{-\int_t^s \rho_u du}+\int_{[t,s)} K_u e^{-\int_u^s \rho_r dr} d\T_u,\quad s\in[t,T+].
\end{equation}
The cost function $J$ represents the total temporary impact costs of the strategy $\T$
on the time interval $[t,T]$ when the initial price impact $D_t=\de$.
Observe that $J$ is well-defined and finite due to Assumption~\ref{basic assumption}.

Let us now define the \emph{value function for continuous trading time}
$U\colon[0,T ]\times[0,\infty)^2\rightarrow[0,\infty)$ as
\begin{equation}
\label{U}
U(t,\de,x):=\inf_{\T \in \cA_t(x)} J(t,\de,\T).
\end{equation}
We also want to discuss \emph{discrete trading time}, i.e. when trading is only allowed at given times
$$0=t_0<t_1<...<t_N=T.$$
Define $\tilde{n}(t):=\inf\{n=0,...,N|t_n \geq t\}$.
We then have to constrain our strategy sets to
\begin{align*}
\cA_t^N&:=\left\{ \T \in \cA_t\,|\,\T_s=0 \text{ on } [t,t_{\tilde{n}(t)}],\,
\T_s=\T_{t_n+} \text{ on } (t_n,t_{n+1}] \text{ for } n=\tilde{n}(t),...,N-1 \right\} \subset \cA_t,\\
\cA_t^N(x)&:=\left \{ \T \in \cA^N_t\,|\,\T_{T+}=x\right\} \subset \cA_t(x),
\end{align*} 
and the \emph{value function for discrete trading time} becomes
\begin{equation}
\label{UN}
U^N(t,\de,x):=\inf_{\T \in \cA_t^N(x)} J(t,\de,\T) \geq U(t,\de,x).
\end{equation}

Note that the optimization problems in continuous time (\ref{U}) and in discrete time (\ref{UN}) only refer to the ask side of the limit order book. The results for optimal trading strategies that we derive in the following sections are hence applicable not only to the specific limit order book model introduced in Section \ref{SecModel}, but also to any model which excludes transaction-triggered price manipulation and where the ask price evolution for pure buying strategies is identical to the ask price evolution in our model. This includes for example models with different depth of the bid and ask sides of the limit order book, or different resiliences of the two sides of the book.

We close this section with the following simple result,
which shows that our problem is economically sensible.

\begin{lemma}[Splitting argument]
\label{split}\mbox{}\\
Doing two separate trades $\xi_\al,\xi_\be >0$ at the same time $s$
has the same effect as trading at once $\xi:=\xi_\al+\xi_\be$, 
i.e., both alternatives incur the same impact costs and the same impact~$D_{s+}$.
\end{lemma}

\begin{proof}
	The impact costs are in both cases
	\begin{eqnarray*}
		\left(D_s+\frac{K_s}{2}\xi \right)\xi&=&D_s \left(\xi_\al+\xi_\be \right)+\frac{K_s}{2}\left(\xi_\al^2+2 \xi_\al \xi_\be 
		+\xi_\be^2\right)\\ 	
		&=&\left(D_s+\frac{K_s}{2} \xi_\al \right)\xi_\al+\left(D_s+K_s \xi_\al+\frac{K_s}{2}\xi_\be \right)\xi_\be
	\end{eqnarray*}
	and the impact $D_{s+}=D_s+K_s \left(\xi_\al+\xi_\be \right)$ after the trade is the same in both cases as well.
\end{proof}
\section{Preparations}\label{Subsec: Preparations}

In this section, we first show that in our model optimal strategies are linear in~$(\delta, x)$, which allows us to reduce the dimensionality of our problem from three dimensions to two dimensions. Thereafter, we introduce the concept of WR-BR structure in Section~\ref{SecBRWR}, which appropriately describes the value function and optimal execution strategies in our model as we will see in Sections~\ref{Subsec: WR-BR deterministic} and~\ref{Subsec: WR-BR deterministic cts}. Finally, we establish some elementary properties of the value function and optimal strategies in Section~\ref{Subsec: CompStat}.

In this entire section, we usually refer only to the continuous time setting, for example, to the value function~$U$. We refer to the discrete time setting only when there is something there to be added explicitly. But all of the statements in this section hold both in continuous time (i.e.~for~$U$) and in discrete time (i.e.~for~$U^N$), and we will later use them in both situations.


\subsection{Dimension reduction of the value function}\label{Subsec: DimReduction}
In this section, we prove a scaling property of the value function which helps us to reduce the dimension of our optimization problem. Our approach exploits both the block shape of the limit order book and the exponential decay of price impact and hence does not generalize easily to more general dynamics of~$D$ as, e.g., in \citeasnoun{PredoiuShaikhetShreve}. We formulate the result for continuous time, although it also holds for discrete time.

\begin{lemma}[Optimal strategies scale linearly]
\label{scaling lemma 1}\mbox{}\\
	For all~$a \in [0,\infty)$ we have
	\begin{equation}\label{U scales Part I}
		U(t,a \de,a x)=a^2 U(t,\de,x).
	\end{equation}
	Furthermore, if~$\T^* \in \cA_t(x)$ is optimal for~$U(t,\de,x)$, then~$a \T^* \in \cA_t(ax)$ is optimal for~$U(t,a \de,ax)$.
\end{lemma}

\begin{proof}
	The assertion is clear for~$a=0$. For any~$a \in (0,\infty)$ and~$\T \in \cA_t$, we get from~(\ref{J}) and~(\ref{D explicit}) that 
	\begin{equation}\label{J scales}
		J(t,a \de,a\T)=a^2 J(t,\de,\T).
	\end{equation}
	Let~$\T^* \in \cA_t(x)$ be optimal for~$U(t,\de,x)$ and~$\bar{\T} \in \cA_t(ax)$ 
	be optimal 	 for~$U(t,a \de,ax)$. If no such optimal strategies exist, the same arguments can be performed with minimizing sequences of strategies. Using~(\ref{J 
	scales}) two times and the optimality of $\T^*,\bar{\T}$, we get
	$$J(t,a \de,\bar{\T}) \leq J(t,a \de,a \T^*)=a^2 J(t,\de,\T^*) 
		\leq  a^2 J \left(t,\de,\frac{1}{a} \bar{\T}\right)=J(t,a \de,\bar{\T}).$$
	Hence, all inequalities are equalities. Therefore,~$a \T^*$ is optimal for~$U(t,a \de,ax)$ and~(\ref{U scales Part I}) holds.
\end{proof}

For~$\de > 0$, we can take~$a=\frac{1}{\de}$ and apply Lemma~\ref{scaling lemma 1} to get
\begin{eqnarray}\label{four to three}
	U(t,\de,x) &=& \de^2 U\left(t,1,\frac{x}{\de} \right)=\de^2 V(t,y) \hspace{.3cm} \text{ with}\\
	\nonumber y &:=& \frac{x}{\de},\\
	\nonumber V(t,y) &:=& U(t,1,y), \hspace{.3cm} V(T,y)=y+\frac{K_T}{2}y^2, \hspace{.3cm} V(t,0) \equiv 0.
\end{eqnarray}
In this way we are able to reduce our three-dimensional value function $U$ defined in~\eqref{U} to a two-dimensional function~$V$. That is $U(t,\de_{fix},x)$ for some $\de_{fix}>0$ or $U(t,\de,x_{fix})$ for some $x_{fix}>0$ already determines the entire value function\footnote{In the following, we will often analyze the function $V$ in order to derive properties of~$U$. Technically this does not directly allow us to draw conclusions for $U(t, 0, x)$, where $\de=0$, since in this case $y=x/\de$ is not defined. The extension of our proofs to the possibility $\de=0$ however is straightforward using continuity arguments (see Proposition~\ref{V cts} below) or alternatively by analyzing $\tilde{V}(t,\tilde{y}):=U(t,\tilde{y},1)$.}. Instead of keeping track of the values $x$ and $\de$ separately, only the ratio of them is important. It should be noted however that the function $V$ itself is not necessarily the value function of a modified optimization problem.
In a similar way we define the function $V^N$ through the function~$U^N$.

\subsection{Introduction to buy and wait regions}\label{SecBRWR}
\begin{figure}[htbp]
 \centering
 \includegraphics[width=0.54\linewidth]{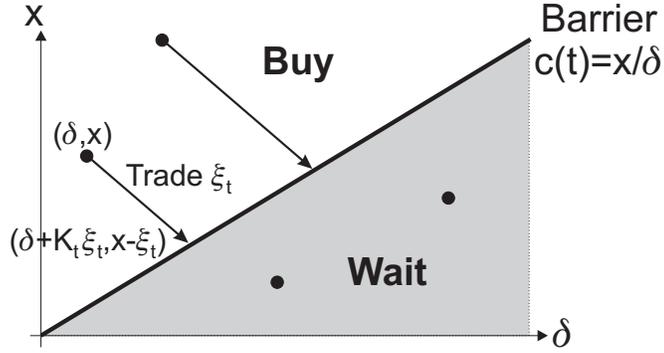} 
 \caption{The~$\delta$-$x$-plane for fixed time~$t$.}
 \label{XD Plane}
\end{figure}

Let us consider an investor who at time~$t$ needs to purchase a position of~$x>0$ in the remaining time until~$T$ and is facing a limit order book dislocated by~$D_t=\delta\ge0$. Any trade~$\xi_t$ at time~$t$ is decreasing the number of shares that are still to be bought, but is increasing~$D$ at the same time (see Figure~\ref{XD Plane} for a graphical representation).
In the~$\de$-$x$-plane, the investor can hence move downwards and to the right. Note that due to the absence of transaction-triggered price manipulation (as shown in Proposition~\ref{only buy orders}) any intermediate sell orders are suboptimal and hence will not be considered. 

Intuitively one might expect the large investor to behave as follows: If there are many shares~$x$ left to be bought and the price deviation~$\de$ is small, then the large investor would buy some shares immediately. In the opposite situation, i.e. small~$x$ and large~$\de$, he would defer trading and wait for a decrease of the price deviation due to resilience. We might hence conjecture that the~$\de$-$x$-plane is divided by a time-dependent barrier into one \emph{buy} region above and one \emph{wait} region below the barrier. Based on the linear scaling of optimal strategies (Lemma~\ref{scaling lemma 1}), we know that if~$(\de,x)$ is in the buy region at time~$t$, then, for any $a>0$, $(a\de, ax)$ is also in the buy region. The barrier between the buy and wait regions therefore has to be a straight line through the origin and the buy and sell region can be characterized in terms of the ratio~$y=\frac{x}{\de}$. In this section, we formally introduce the buy and wait regions and the barrier function. In Sections~\ref{Subsec: WR-BR deterministic} and~\ref{Subsec: WR-BR deterministic cts}, we prove that such a barrier exists for discrete and continuous trading time respectively. In contrast to the case of a time-varying but deterministic illiquidity~$K$ considered in this paper, for \emph{stochastic~$K$}, this barrier conjecture holds true in many, but not all cases, see~\citeasnoun{Diss}.

We first define the buy and wait regions and subsequently define the barrier function. Based on the above scaling argument, we can limit our attention to points~$(1,y)$ where~$\de=1$, since for a point~$(\de, x)$ with~$\de>0$ we can instead consider the point~$(1, x/\de)$. 

\begin{definition}[Buy and wait region]
\label{BR definition}\mbox{}\\
	For any~$t \in[0,T]$, we define the \emph{inner buy region} as
	$$Br_t:=\left\{y \in (0,\infty) \,|\, \exists\xi \in (0,y)\colon U(t,1,y)=U \left (t,1+K_t \xi,y-\xi \right)+\left(1+\frac{K_t}{2} \xi \right) \xi 
	\right\},$$
	and call the following sets the \emph{buy region} and \emph{wait region} at time~$t$:
	$$BR_t:=\overline{Br_t}, \hspace{.5 cm} WR_t:=[0,\infty)\setminus Br_t$$
	(the bar means closure in~$\R$).
\end{definition}

The inner buy region at time $t$ hence consists of all values $y$ such that immediate buying at the state $(1,y)$ is value preserving. The wait region on the other hand contains all values $y$ such that any non-zero purchase at $(1,y)$ destroys value.
Let us note that $Br_T=(0,\infty)$, $BR_T=[0,\infty)$ and $WR_T=\{0\}$.

Regarding Definition~\ref{BR definition}, the following comment is in order.
We do not claim in this definition that $Br_t$ is an open set.
A~priori one might imagine, say, the set $(10,20]$ as the inner buy region at some time point.
But what we can say from the outset is that, due to the splitting argument (see~Lemma~\ref{split}),
$Br_t$ is in any case a union of (not necessarily open) intervals or the empty set.

The wait-region/buy-region conjecture can now be formalized as follows.

\begin{definition}[WR-BR structure]
\label{WR-BR structure definition}\mbox{}\\
	The value function $U$ has \emph{WR-BR structure} if there exists a \emph{barrier function} 
	$$c\colon[0,T]  \rightarrow [0,\infty]$$
	such that for all~$t \in[0,T]$,
	$$Br_t=\left(c(t),\infty \right)$$
	with the convention~$(\infty,\infty):=\emptyset$. For the value function~$U^N$ in discrete time to have WR-BR structure, we only consider~$t \in \{t_0,...,t_N \}$ 
	and set~$c^N(t)=\infty$ for~$t \notin \{t_0,...,t_N \}$. 
\end{definition}

\begin{figure}[htbp]
 \centering
 \includegraphics[width=0.54\linewidth]{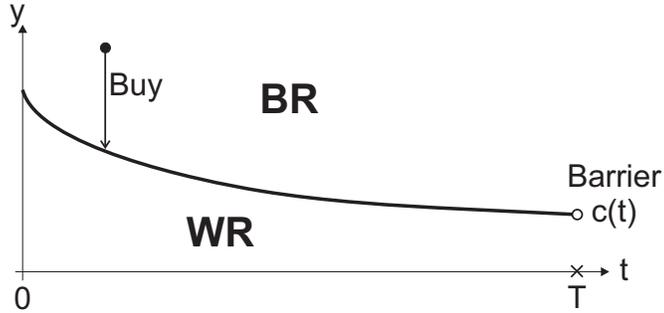} 
 \caption{Schematic illustration of the buy and wait regions in continuous time. 
 }
 \label{Barrier figure}
\end{figure}

Let us note that we always have $c(T)=0$.
Below we will see that it is indeed possible to have $c(t)=\infty$,
i.e. at time $t$ any strictly positive trade is suboptimal no matter at which state we start.
For~$c(t)< \infty$, having WR-BR structure means that $BR_t \cap WR_t=\{c(t)\}$.
Figure~\ref{Barrier figure} illustrates the situation in continuous time.

Thus, up to now we have the following intuition.
An optimal strategy is suggested by the barrier function whenever the value function has WR-BR structure. If the position of the large investor at time~$t$ satisfies~$\frac{x}{\de} > c(t)$, then the portfolio is in the buy region. We then expect that it is optimal to execute the largest discrete trade $\xi \in (0,x)$ such that the new ratio of remaining shares over price deviation $\frac{x-\xi}{\de+K_t \xi}$ is still in the buy region, i.e. the optimal trade is
$$\xi^* = \frac{x-c(t) \de}{1+K_t c(t)},$$
which is equivalent to
$$c(t) = \frac{x-\xi^*}{\de+K_t \xi^*}.$$
Notice that the ratio term $\frac{x-\xi}{\de+K_t \xi}$ is strictly decreasing in~$\xi$. Consequently, trades have the effect of reducing the ratio as indicated in Figure~\ref{Barrier figure}, while the resilience effect increases it. That is one trades just enough shares to keep the ratio~$y$ below the barrier.\footnote{Intuitively, this implies that apart from a possible initial and final impulse trade, optimal buying occurs in in\-fi\-ni\-te\-si\-mal amounts provided that $c$ is continuous in~$t$ on~$[0,T)$. For diffusive~$K$ as in \citeasnoun{Diss}, this would lead to singular optimal controls.}

In Figure~\ref{Barrier figure} we demonstrate an intuitive case where the barrier decreases over time, i.e. buying becomes more aggressive as the investor runs out of time. This intuitive feature however does not need to hold for all possible evolutions of $K$ and $\rho$ as we will see e.g. in Figure~\ref{Discrete figure}.

Below we will see that the intuition presented above always works in discrete time:
namely, the value function $U^N$ always has WR-BR structure,
there exists a unique optimal strategy, which is of the type
``trade to the barrier when the ratio is in the buy region, do not trade when it is in the wait region''
(see Section~\ref{Subsec: WR-BR deterministic}).
In continuous time the situation is more delicate.
It may happen, for example, that the value function $U$ has WR-BR structure,
but the strategy consisting in trading towards the barrier is not optimal
(see the example in the beginning of Section~\ref{Subsec: WR-BR deterministic cts},
where an optimal strategy does not exist).
However, if the illiquidity~$K$ is continuous, there exists an optimal strategy,
and, under additional technical assumptions, it is unique
(see Section~\ref{Subsec: WR-BR deterministic cts}).
Moreover, if $K$ and $\rho$ are smooth and satisfy some further technical conditions,
we have explicit formulas for the barrier and for the optimal strategy
(see Section~\ref{Subsec: Euler Lagrange}).


\subsection{Some properties of the value function and buy and wait regions}\label{Subsec: CompStat}
We first state comparative statics satisfied by both the continuous and the discrete time value function. The value function is increasing in~$t,\de,x$ and the price impact coefficient~$K$ as well as decreasing with respect to the resilience speed function.

\begin{proposition}[Comparative statics for the value function]
\label{comparative statics for U}\mbox{}

\vspace{-4mm}
	\begin{enumerate}[a)]
		\item The value function is nondecreasing in~$t,\de,x$.
		\item Fix~$t \in [0,T]$. Assume that~$0<\check{K}_s \leq \hat{K}_s$ for all~$s \in [t,T]$. Then the value function corresponding 
	  			to~$\check{K}$ is less than or equal to the one corresponding to~$\hat{K}$.
	  \item Fix~$t \in [0,T]$. Assume that~$0<\check{\rho}_s \leq \hat{\rho}_s$ for all~$s \in [t,T]$. Then the value function corresponding 
	  			to~$\hat{\rho}$ is less than or equal to the one corresponding to~$\check{\rho}$.
	\end{enumerate}
\end{proposition}

The proof is straightforward.

\begin{proposition}[Continuity of the value function]
\label{V cts}\mbox{}\\
For each $t\in[0,T]$, the functions
$$U(t,\cdot,\cdot)\colon[0,\infty)^2\to[0,\infty)
\quad\text{and}\quad
V(t,\cdot)\colon[0,\infty)\to[0,\infty)$$
are continuous.
\end{proposition}

\begin{proof}
Due to Lemma~\ref{scaling lemma 1} it is enough to prove that the function $U(t,\cdot,\cdot)$ is continuous.
Let us fix $t\in[0,T]$, $x\ge0$, $0\le\de_1<\de_2$, $\ep>0$ and take a strategy $\T^\ep\in\cA_t(x)$ such that
$$
J(t,\de_1,\T^\ep)<U(t,\de_1,x)+\ep.
$$
For~$i=1,2$, we define
$$
D^i_s:=\de_i e^{-\int_t^s \rho_u du}+\int_{[t,s)} K_u e^{-\int_u^s \rho_r dr} d\T^\ep_u,\quad s\in[t,T+].
$$
Using Proposition~\ref{comparative statics for U}, we get
\begin{align*}
U(t,\de_1,x)
\le U(t,\de_2,x)
&\le\int_{[t,T]}\left(D^2_s+\frac{K_s}2\De\T^\ep_s\right)\,d\T^\ep_s\\
&\le\int_{[t,T]}\left(D^1_s+\frac{K_s}2\De\T^\ep_s\right)\,d\T^\ep_s+(\de_2-\de_1)x\\
&=J(t,\de_1,\T^\ep)+(\de_2-\de_1)x
<U(t,\de_1,x)+\ep+(\de_2-\de_1)x.
\end{align*}
Thus, for each fixed $t\in[0,T]$ and $x\ge0$, the function $U(t,\cdot,x)$ is continuous on $[0,\infty)$.
For $t\in[0,T]$, $\de\ge0$ and $x>0$, by Lemma~\ref{scaling lemma 1}, we have
$$
U(t,\de,x)=x^2U(t,\de/x,1),
$$
hence the function $U(t,\cdot,\cdot)$ is continuous on $[0,\infty)\times(0,\infty)$.
Considering the strategy of buying the whole position~$x$ at time~$t$, we get
$$
U(t,\de,x)\le\left(\de+\frac{K_t}2x\right)x
\xrightarrow[x\searrow0]{}0=U(t,\de,0),
$$
i.e. the function $U(t,\cdot,\cdot)$ is also continuous on $[0,\infty)\times\{0\}$.
This concludes the proof.
\end{proof}

\begin{proposition}[Trading never completes early]
\label{lemma no finish early}\mbox{}\\
	For all~$t\in[0,T)$, $\de \in [0,\infty)$ and~$x\in (0,\infty)$, the value function satisfies
	$$U(t,\de,x)<\left(\de+\frac{K_t}{2}x \right)x,$$
	i.e. it is never optimal to buy the whole remaining position at any time $t\in[0,T)$.
\end{proposition}

\begin{proof}
	For~$\ep \in [0,x]$, define the strategies~$\T^\ep \in \cA_t(x)$ that buy~$(x-\ep)$ shares at~$t$ and~$\ep$ shares at~$T$. 
	The corresponding costs are
	$$J \left(t,\de,\T^\ep \right)=\left(\de+\frac{K_t}{2}(x-\ep)\right)(x-\ep)+\left(\left(\de+K_t[x-\ep]\right)e^{-\int_t^T \rho_s 
	ds}+\frac{K_T}{2}\ep \right)\ep.$$
	Clearly,
	$$U(t,\de,x) \leq J \left(t,\de,\T^0 \right)= \left(\de+\frac{K_t}{2}x \right)x,$$
	but we never have equality since
	$$\frac{\partial}{\partial \ep}J \left(t,\de,\T^\ep \right) \Big |_{\ep=0}=- \left(1-e^{-\int_t^T \rho_s ds} \right)(K_t x+\de)<0.$$
\end{proof}

As discussed above, we always have $Br_T=(0,\infty)$ and $WR_T=\{0\}$.
In two following propositions we discuss $Br_t$ (equivalently,~$WR_t$) for $t\in[0,T)$.

\begin{proposition}[Wait region near~$0$]
\label{prop:WR near 0}\mbox{}\\
Assume that the value function $U$ has WR-BR structure with the barrier~$c$.
Then for any $t\in[0,T)$, $c(t)\in(0,\infty]$
(equivalently, there exists $\ep>0$ such that $[0,\ep)\subset WR_t$).
\end{proposition}

\begin{proof}
We need to exclude the possibility $c(t)=0$, i.e. $Br_t=(0,\infty)$.
But if $Br_t=(0,\infty)$, we get by Proposition~\ref{V cts}
that for any $y>0$,
$$
V(t,y)=\left(1+\frac{K_t}2y\right)y,
$$
which contradicts Proposition~\ref{lemma no finish early}.
\end{proof}

The following result illustrates that the barrier can be infinite.

\begin{proposition}[Infinite barrier]
\label{lemma infinite barrier}\mbox{}\\
	Assume there exist~$0\leq t_1<t_2\leq T$ such that
	$$K_s e^{-\int_s^{t_2} \rho_u du} >K_{t_2}\quad\text{for all }s \in [t_1,t_2).$$
	Then $Br_s= \emptyset$ for~$s \in [t_1,t_2)$.
\end{proposition}

In particular, if the assumption of Proposition~\ref{lemma infinite barrier} holds
with $t_1=0$ and $t_2=T$, then the value function has WR-BR structure
and the barrier is infinite except at terminal time~$T$.

\begin{proof}
	For any~$s \in [t_1,t_2)$,~$\de \in [0,\infty)$, $x \in (0,\infty)$ and~$\T \in \cA_s(x)$ with~$\T_{t_2}>0$, we get the following by 
	applying~(\ref{D explicit}), the assumption of the proposition, monotonicity of~$J$ in~$\de$, and integration by parts as in~\eqref{x square in by parts}
	\begin{eqnarray*}
		J\left(s,\de,\T \right) &=& \int_{[s,t_2)} \left(D_u+\frac{K_u}{2} \De \T_u \right) d \T_u + J\left(t_2,D_{t_2},(\T_u-\T_{t_2})_{u\in [t_2,T+]} \right)\\
		&\geq& \int_{[s,t_2)} \left(\de e^{-\int_{s}^u \rho_r dr} +\int_{[s,u)} K_r e^{-\int_{r}^u \rho_w dw}d\T_r +\frac{K_u}{2} \De \T_u\right)d\T_u\\
		&& + J\left(t_2,\de e^{-\int_{s}^{t_2} \rho_u du}+K_{t_2} \T_{t_2},(\T_u-\T_{t_2})_{u\in [t_2,T+]} \right)\\
		&>& \left(\de e^{-\int_{s}^{t_2} \rho_u du}+ \frac{K_{t_2}}{2} \T_{t_2}\right) \T_{t_2} +J\left(t_2,\de e^{-\int_{s}^{t_2} \rho_u 
		du}+K_{t_2} \T_{t_2},(\T_u-\T_{t_2})_{u\in [t_2,T+]} \right).
	\end{eqnarray*}
	That is it is strictly suboptimal to trade on~$[s,t_2)$. In particular,~$Br_s=\emptyset$.
\end{proof}

Proposition~\ref{lemma infinite barrier} can be extended in the following way.

\begin{proposition}[Infinite barrier, extended version]
\label{lemma infinite barrier extended}\mbox{}\\
	Let~$K$ be continuous and assume there exist~$0\leq t_1<t_2\leq T$ such that
	$$K_{t_1} e^{-\int_{t_1}^{t_2} \rho_u du} >K_{t_2}.$$
	Then $Br_{t_1}= \emptyset$.
\end{proposition}

\begin{proof}
	Define~$\tilde{t}$ as the minimal value of the set
	$$\underset{t \in [t_1,t_2]}{\operatorname{argmin}} \, K_t e^{\int_0^t \rho_u du}$$
	with~$\tilde{t}$ being well-defined due to the continuity of~$K$.
	Then we know that~$\tilde{t} > t_1$. By definition of~$\tilde{t}$, we have that for all~$t\in [t_1,\tilde{t})$
	$$K_t e^{\int_0^t \rho_u du}	 > K_{\tilde{t}} e^{\int_0^{\tilde{t}} \rho_u du}$$
	and hence
	$$K_t e^{-\int_t^{\tilde{t}} \rho_u du}	 > K_{\tilde{t}}.$$
	By Proposition~\ref{lemma infinite barrier}, we can conclude that~$Br_t= \emptyset$ for all~$t\in [t_1,\tilde{t})$ and hence in particular for~$t=t_1$.
\end{proof}

\section{Discrete time}\label{Subsec: WR-BR deterministic}
In this section we show that the optimal execution problem in discrete time has WR-BR structure. Let us first rephrase the problem in the discrete time setting and define~$K_n:=K_{t_n}$, $D_n:=D_{t_n}$ and~$\xi_n:=\De \T_{t_n}$ for~$n=0,...,N$. The optimization problem~(\ref{U}) can then be expressed as 
\begin{equation}\label{EqnValueFnDiscreteTime}
U^N(t_n,\de,x)=\inf_{\stackrel{\xi_j \in [0,x]}{\sum \xi_j=x}} \sum_{j=n}^N \left( D_{j}+\frac{K_j}{2} \xi_j \right) \xi_j.
\end{equation}
with~$D_{n}=\de$ and~$D_{j+1}=(D_{j}+K_j \xi_j)a_j$, where
\begin{equation}\label{a_j}
	a_j:=\exp\left(- \int_{t_j}^{t_{j+1}} \rho_s ds\right).
\end{equation}
Recall the dimension reduction from Lemma~\ref{scaling lemma 1}
$$U^N(t_n,\de,x)=\de^2 V^N\left(t_n,\frac{x}{\de} \right) \: \text{ with } \: V^N(t_n,y):=U^N(t_n,1,y).$$
The following theorem establishes the WR-BR structure in discrete time. 

\begin{theorem}[Discrete time: WR-BR structure]
\label{prop deterministic WR-BR}\mbox{}\\
The discrete time value function $U^N$ has WR-BR structure with some barrier function~$c^N$.
There exists a unique optimal strategy, which corresponds to the barrier~$c^N$ as described in Section~\ref{SecBRWR}.
Furthermore, $V^N(t_n,\cdot):[0,\infty) \rightarrow [0,\infty)$ has the following properties for~$n=0,...,N$.

	\vspace{-4mm}
	\begin{enumerate}[(i)]
		\item It is {\bf continuously differentiable}.
		\item It is {\bf piecewise quadratic}, i.e.,~there exists~$M \in \N$, constants~$(\al_i,\be_i,\ga_i)_{i=1,...,M}$	and~$0<y_1<y_2<...<y_M=\infty$ 
		such that
		$$V^N(t_n,y)=\al_{m(y)}y^2+\be_{m(y)}  y+\ga_{m(y)}$$
		for the index function~$m:[0,\infty) \rightarrow \{1,...,M\}$ with~$m(y):=\min 	\{i|y \leq y_i\}$.
		\item The coefficients~$(\al_i,\be_i,\ga_i)_{i=1,...,M}$ from (ii) satisfy the {\bf inequalities}
		\begin{eqnarray}\label{alpha inequalities}
			\al_i, \, \be_i & > & 0,\\
			\nonumber 4 \al_i \ga_i+\be_i-\be_i^2 & \geq & 0,\\
			\nonumber y_{i-1} \be_i+2 \ga_i & \geq & 0.
		\end{eqnarray}
	\end{enumerate}
\end{theorem}
 
The properties (i)--(iii) of~$V^N$ included in the above theorem will be exploited in the backward induction proof of the WR-BR structure. The piecewise quadratic nature of the value function occurs, since the price impact~$D$ is affine in the trade size and is multiplied by the trade size in the value function~\eqref{EqnValueFnDiscreteTime}. Let us, however, note that the value function in continuous time is no longer piecewise quadratic.

We prove Theorem~\ref{prop deterministic WR-BR} by backward induction. As a preparation we investigate the relationship of the function~$V^N$ at times~$t_{n}$ and~$t_{n+1}$. They are linked by the dynamic programming principle:
\begin{eqnarray}\label{U^N}	
       \nonumber V^N(t_n,y) &=& \min_{\xi \in [0,y]} \left\{ \left(1+\frac{K_n}{2}\xi \right)\xi+U^N \left(t_{n+1},(1+K_n \xi)a_n,y-\xi \right) \right\}\\
       &=& \min_{\xi \in [0,y]} \left\{ \left(1+\frac{K_n}{2}\xi \right)\xi+(1+K_n \xi)^2 a^2_n V^N \left(t_{n+1},\frac{y-\xi}{(1+K_n \xi)a_n} \right) \right\}.
\end{eqnarray}
Instead of focusing on the optimal trade~$\xi$, we can alternatively look for the optimal new ratio~$\eta(\xi):=\frac{y-\xi}{1+K_n \xi}$ of remaining shares over price deviation. Note that~$\eta$ is decreasing in the trade size~$\xi$ and bounded between zero (if the entire position is traded at once) and the current ratio~$y$ (if nothing is traded). A straightforward calculation confirms that~(\ref{U^N}) is equivalent to 
    \begin{equation}\label{optimal barrier eq}
	  		V^N(t_n,y) =  \frac{1}{2K_n}\left[(1+K_n y)^2 \min_{\eta \in [0,y]} L^N(t_n,\eta)-1 \right], 
	  \end{equation}
	  where
		\begin{equation}\label{def L}
		   L^N(t_n,\eta) := \frac{1+2 K_n a_n^2 V^N(t_{n+1},\eta a_n^{-1})}{(1+K_n \eta)^2}.
		\end{equation}
Note that in~(\ref{optimal barrier eq}) the minimization is taken over~$\eta$ instead of~$\xi$. Furthermore, the function~$L^N$ depends on~$\eta$, but not on~$y$ or~$\xi$ separately. In the sequel, the function~$L^N$ will be essential in several arguments. The following lemma will be used in the proof of Theorem~\ref{prop deterministic WR-BR}.

\begin{lemma}\label{lemma L}
	Let~$a \in (0,1),\ka>0$ and let the function~$v:[0,\infty) \rightarrow [0,\infty)$ satisfy~(i), (ii), (iii) given 
	in Theorem~\ref{prop deterministic WR-BR}. Then the following statements hold true.
	
\vspace{-4mm}
	\begin{enumerate}[(a)]
	\item There exists~$c^* \in [0,\infty]$ such that
	$$L(y):=\frac{1+2 \ka a^{2}v(y a^{-1})}{(1+\ka y)^2}, \hspace{.2cm} y \in [0,\infty),$$
	is strictly decreasing for~$y \in [0,c^*)$ and strictly increasing for~$y \in (c^*,\infty)$.
	\item The function
	$$\tilde{v}(y):=\left\{ \begin{array}{cl} \frac{1}{2 \ka}\left[(1+\ka y)^2 L(c^*)-1 \right] & \text{if } y>c^*\\ a^{2 }v(y a^{-1}) 
	&	\text{otherwise} \end{array}\right  \}$$
	again satisfies~(i), (ii), (iii) with possibly different coefficients.
	\end{enumerate}
\end{lemma}

\begin{proof}[Proof of Theorem~\ref{prop deterministic WR-BR}]
	We proceed by backward induction. Notice that~$V^N(t_N,y)=\left(1+\frac{K_N}{2}y \right)y$ fulfills~(i), (ii), (iii) 
	with~$M=1,\al_1=\frac{K_N}{2},\be_1=1,\ga_1=0$. 
	Let us consider the induction step from~$t_{n+1}$ to~$t_n$. We are going to use Lemma~\ref{lemma L} 
	for~$a=a_n,\ka=K_n,v=V^N(t_{n+1},\cdot)$. We then have that~$L=L^N(t_n,\cdot)$ and we obtain~$c^*$ as the unique minimum of~$L^N(t_n,\cdot)$ from Lemma~\ref{lemma L}~(a). From~(\ref{optimal barrier eq}) we see that the unique optimal value for~$\eta$ is given by  
	$$\eta^*:=\underset{\eta \in [0,y]}{\operatorname{argmin}} \frac{1}{2K_n}\left[(1+K_n y)^2 L^N(t_n,\eta)-1 \right]=\min 
		\left\{y,c^* \right\}$$
	and accordingly that the unique optimal trade is given by
	$$\xi^*:=\xi \left(\eta^*\right)=\max \left\{0,\frac{y-c_n }{1+K_n c_n} \right\}.$$
	Therefore we have a unique optimal strategy and the value function has WR-BR structure with~$c^N(t_n):=c^*$.
	Plugging~$\xi^*$ into~(\ref{U^N}) and applying the definition of~$V^N$ yields~$V^N(t_n,y)=\tilde{v}(y)$. Lemma~\ref{lemma L}~(b) now concludes the induction step. 
\end{proof}

\begin{proof}[Proof of Lemma~\ref{lemma L}]
	(a) The function~$L$ is continuously differentiable with
	\begin{eqnarray}\label{def L derivative}
		L'(y)&=&  \frac{2 \ka}{(1+\ka y)^3} l(y),\\
		\nonumber l(y) &:=& y \left(2 \al_{m(y a^{-1})}-\ka \beta_{m(y a^{-1})} a \right) + \left(\be_{m(y a^{-1})} a-2 \ka \ga_{m(y a^{-1})} a^{2}-1 \right).
	\end{eqnarray}
	First of all, we show that there is no interval where~$L$ is constant. Assume there would be an interval where~$l$ 
	is zero, i.e.,~there exists~$i \in 
	\{1,...,M\}$ such that~$(2 \al_i-\ka \beta_i a)=0$ and~$(\be_i a-2 \ka \ga_i a^{2}-1)=0$. Solving these equations for~$\al$ 
	respectively~$\ga$ yields
	$$4 \al_i \ga_i+a^{-1} \be_i-\be_i^2=0.$$
	This is a contradiction to~(\ref{alpha inequalities}).
	
	Let us assume $l(\check{y})>0$ for some~$\check{y} \in[0,\infty)$ with~$j:=m(\check{y} a^{-1})$. We are done if we can conclude 
	$l(\hat{y})>0 \text{ for all } \hat{y} \in [\check{y},\infty)$.	Because of the continuity of~$l$, it is sufficient to show that~$L$ keeps 
	increasing on~$[\check{y},y_j]$, i.e.,~we need to show~$l(\hat{y})>0 \text{ for all } \hat{y} \in [\check{y},y_j]$. Due to the form of~$l$, this is 
	guaranteed when~$2 \al_j-\ka \be_j a>0$. Let us suppose that this term would be negative which is equivalent to~$2 \al_j	\be^{-1}_j a^{-1} \leq 
	\ka$. Together with the inequalities from~(\ref{alpha inequalities}) one gets
	\begin{eqnarray*}
		a l(\check{y}) &=&	-  \ka \, a \left(\check{y} a^{-1} \be_j+2 \ga_j \right)+ \left(2 \check{y} a^{-1} 	\al_j+\be_j-a^{-1} \right)\\
		& \leq& -2 \al_j \be^{-1}_j \left(\check{y} a^{-1} \be_j+2 \ga_j \right)+ \left(2 \check{y} a^{-1} \al_j+\be_j-a^{-1} \right)\\
		&=& -\frac{1}{\be_j} \left(4 \al_j \ga_j+\be_j a^{-1}-\be_j^2 \right) < 0.
	\end{eqnarray*}
	This is a contradiction to~$l(\check{y})>0$.
	
	(b) If~$c^*$ is finite, the function~$\tilde{v}$ is continuously differentiable at~$c^*$ since a brief calculation shows 
	that~$\tilde{v}'(c^*-)=\tilde{v}'(c^*+)$ is equivalent to~$l(c^*)=0$. We have
	$$\tilde{v}(y)=\tilde{\al}_{\tilde{m}(y)} y^2+\tilde{\be}_{\tilde{m}(y)} y+\tilde{\ga}_{\tilde{m}(y)},$$
	i.e. $\tilde{v}$ is piecewise quadratic with~$\tilde{M}=1+m(c^* a^{-1})$, 
	$\tilde{y}_{\tilde{M}-1}:=c^*$, $\tilde{y}_i:=y_i a$ 
	for~$i=1,...,\tilde{M}-2$ and
	\begin{eqnarray}\label{coefficient update}
		\tilde{\al}_{\tilde{M}} &=& \frac{\ka}{2}L(c^*)>0, \: \tilde{\be}_{\tilde{M}}=L(c^*)>0, \: \tilde{\ga}_{\tilde{M}}=\frac{L(c^*)-1}{2 \ka},\\
		\nonumber \tilde{\al}_i &=& \al_i>0, \: \tilde{\be}_i=a \be_i>0, \: \tilde{\ga}_i=a^{2} \ga_i \: \text{ for } i=1,...,\tilde{M}-1.
	\end{eqnarray}
	We therefore get
	$$4 \tilde{\al}_i \tilde{\ga}_i+\tilde{\be}_i-\tilde{\be}_i^2 =\left\{ \begin{array}{cl} 0 & \text{if } i=\tilde{M}\\ a^{2} \left(4 \al_i 
	\ga_i+a^{-1} \be_i-\be_i^2 \right) &	\text{otherwise} \end{array}\right  \} \geq 0.$$
	It remains to show that~$\tilde{v}$ also inherits the last inequality in~(\ref{alpha inequalities}) from~$v$. For~$y \leq c^*$,
	$$y \tilde{\be}_{\tilde{m}(y)}+2 \tilde{\ga}_{\tilde{m}(y)}=a^{2 } \left(y a^{-1} \be_{m(y a^{-1})}+2  \ga_{m(y a^{-1})} \right) 
	\geq 0.$$
	Due to~$\tilde{v}$ being continuously differentiable in~$c^*$, we get
	\begin{eqnarray*}
	\tilde{\al}_{\tilde{M}} (c^*)^2+\tilde{\be}_{\tilde{M}} c^*+\tilde{\ga}_{\tilde{M}} &=& 
	\tilde{\al}_{\tilde{M}-1} (c^*)^2+\tilde{\be}_{\tilde{M}-1} c^*+\tilde{\ga}_{\tilde{M}-1} ,\\
	2 \tilde{\al}_{\tilde{M}} c^*+\tilde{\be}_{\tilde{M}} &=& 2 \tilde{\al}_{\tilde{M}-1} c^*+\tilde{\be}_{\tilde{M}-1}.
	\end{eqnarray*}
	Taking two times the first equation and subtracting~$c^*$ times the second equation yields
	$$c^* \tilde{\be}_{\tilde{M}}+2 \tilde{\ga}_{\tilde{M}}=c^* \tilde{\be}_{\tilde{M}-1}+2 \tilde{\ga}_{\tilde{M}-1}.$$
	Since we already know that the right-hand side is positive, also~$y \tilde{\be}_{\tilde{M}}+2 \tilde{\ga}_{\tilde{M}} \geq 0$ for 
	all~$y>c^*$.
\end{proof}

We need the following lemma as a preparation for the WR-BR proof in continuous time.

\begin{lemma}\label{lemma deterministic cts WRBR 2}
	Let~$K$ be continuous. Then at least one of two following statements is true:
	
	\vspace{-4mm}
	\begin{itemize}
		\item The function~$y \mapsto L^N(0,y)$ is convex on~$\left[0,c^N(0) \right)$;
		\item The continuous time buy region is simply~$Br_0= \emptyset$, i.e.~$c(0)=\infty$.
	\end{itemize}
\end{lemma}

We stress that the first statement in this lemma concerns discrete time,
while the second one concerns the continuous time optimization problem.

\begin{proof}
	Recall that the definition of~$L^N(0,\cdot)$ from~(\ref{def L}) contains~$V^N(t_1,\cdot)$ which is continuously differentiable 
	and piecewise quadratic with coefficients~$(\al_i,\be_i,\ga_i)$. Analogously to~(\ref{def L derivative}), it turns out that
	\begin{eqnarray*}
		\frac{\partial}{\partial y} L^N(0,y)&=&\frac{2 K_0}{(1+K_0 y)^3} \Bigg[y \left(2 \al_{m\left(y e^{\int_0^{t_1}\rho_s ds}\right)}- K_0 
		\be_{m\left(y	e^{\int_0^{t_1}\rho_s ds}\right)} e^{-\int_0^{t_1}\rho_s ds} \right) \\
		&&+ \left(\be_{m\left(y e^{\int_0^{t_1}\rho_s ds}\right)} e^{-\int_0^{t_1}\rho_s ds}+2K_0 \ga_{m\left(y e^{\int_0^{t_1}\rho_s 
		ds}\right)}	e^{-2\int_0^{t_1}\rho_s ds} -1\right) \Bigg].
	\end{eqnarray*}
	We distinguish between two cases. First assume that all~$i$ satisfy~$( 2 \al_i-K_0 \be_i e^{-\int_0^{t_1}\rho_s ds}) \geq 0$. 
	Then~$\frac{\partial}{\partial y} L^N(0,\cdot)$ must be increasing on~$[0,c^N(0))$ as desired, since~$L^N(0,\cdot)$ is decreasing on 
	this interval as we know from Lemma~\ref{lemma L}.
	
	Assume to the contrary that there exists~$i$ such that~$(2 \al_i-K_0 \be_i e^{-\int_0^{t_1}\rho_s ds})< 0$. Recall how~$\al_i$ and~$\be_i$ are 
	actually computed in the backward induction of Theorem~\ref{prop deterministic WR-BR}. In each induction step, Lemma~\ref{lemma L} is used and 
	the coefficients~$\tilde{\al}_{\tilde{M}},\tilde{\be}_{\tilde{M}}$ get updated in~(\ref{coefficient update}). It gets clear that there exists~$n \in 
	\{1,...,N\}$ such that
	$$2 \al_i-K_0 \be_i e^{-\int_0^{t_1}\rho_s ds} = \left(K_{t_n}-K_0 e^{-\int_0^{t_n}\rho_s ds} \right) L^N\left(t_n,c^N(t_n) \right).$$
	We get the resilience multiplier~$e^{-\int_0^{t_n}\rho_s ds}$ thanks to the adjustment~$\tilde{\be}_i=a \be_i$ from the second line 
	of~(\ref{coefficient update}). Due to~$L^N$ being positive, it follows that
	$$K_{t_n} < K_0 e^{-\int_0^{t_n}\rho_s ds}.$$ 
	That is for this choice of~$K$, it cannot be optimal to trade at~$t=0$ as we see from Proposition~\ref{lemma infinite barrier extended}.
	Hence, the buy region at~$t=0$ is the empty set for both discrete and continuous time.
\end{proof}

The proof of Theorem~\ref{prop deterministic WR-BR} is constructive. It not only establishes the existence of a unique barrier, but also provides means to calculate the barrier numerically through the following recursive algorithm. 

\fbox{ \begin{minipage}{14.95cm}
Initialize value function~$V^N(t_N,y)=\left(1+\frac{K_N}{2}y \right)y$\\
	For~$n=N-1,...,0$\\
		\hspace*{1 cm} Set~$L^N(t_n,y):=\frac{1+2 K_n a_n^{2} V^N\left(t_{n+1},y a_n^{-1} \right)}{\left(1+K_n y \right)^2}$\\
		\hspace*{1 cm} Compute~$c^N(t_n):=c_n:=\underset{y \geq 0}{\operatorname{argmin}} \hspace{.1 cm} L^N(t_n,y)$\\
		\hspace*{1 cm} Set~$V^N(t_n,y):=\left\{ \begin{array}{cl} \frac{1}{2 K_n}\left[(1+K_n y)^2 L^N(t_n,c_n)-1 \right] & \text{if } y>c_n\\ 
		a_n^{2}V^N(t_{n+1},y a_n^{-1}) &		\text{otherwise} \end{array}\right  \}$
\end{minipage} }

\begin{figure}[htbp]
 \centering
 \includegraphics[width=0.49\linewidth]{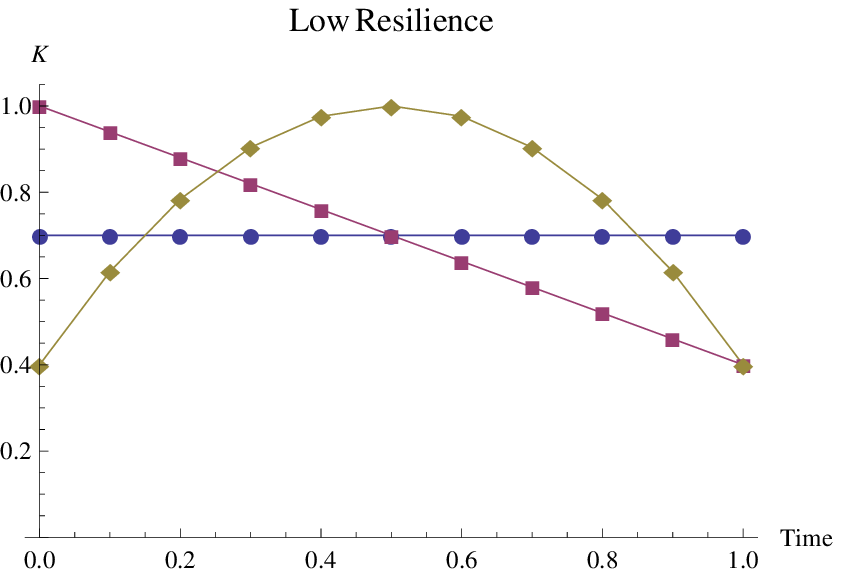}
 \includegraphics[width=0.49\linewidth]{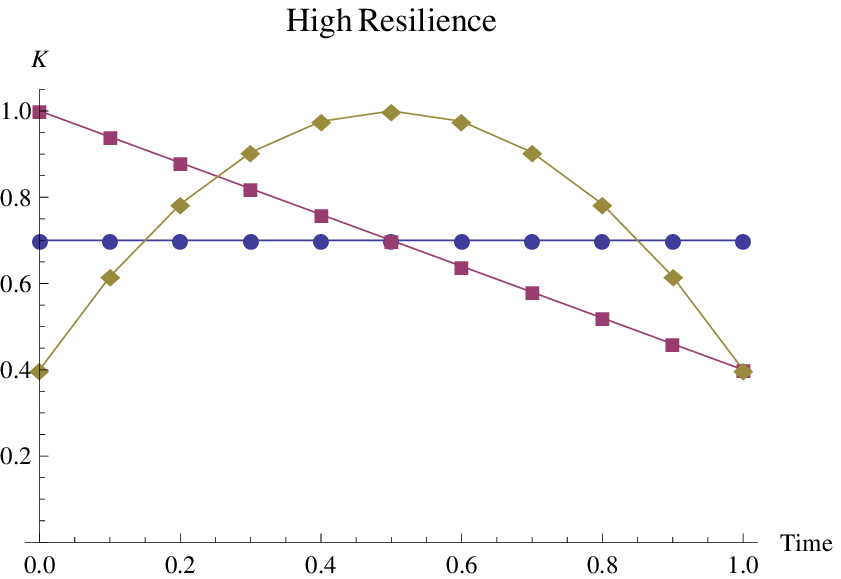}
 \includegraphics[width=0.49\linewidth]{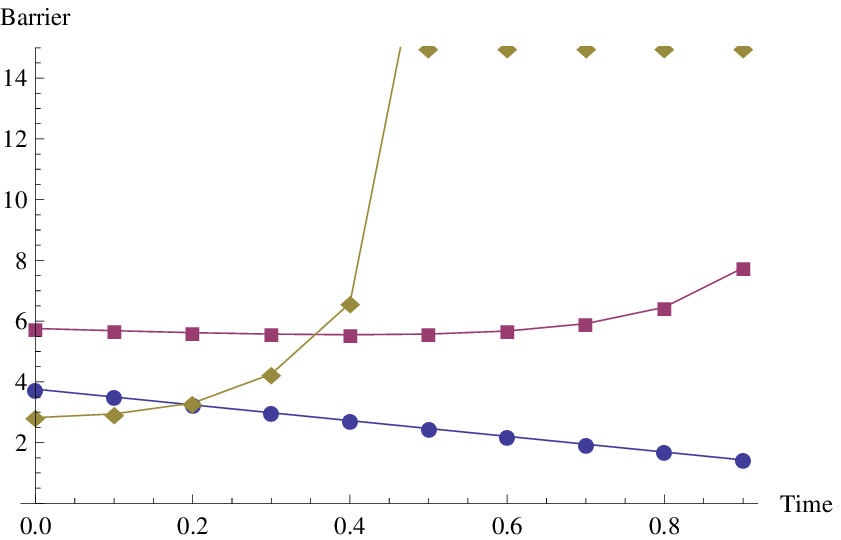} 
 \includegraphics[width=0.49\linewidth]{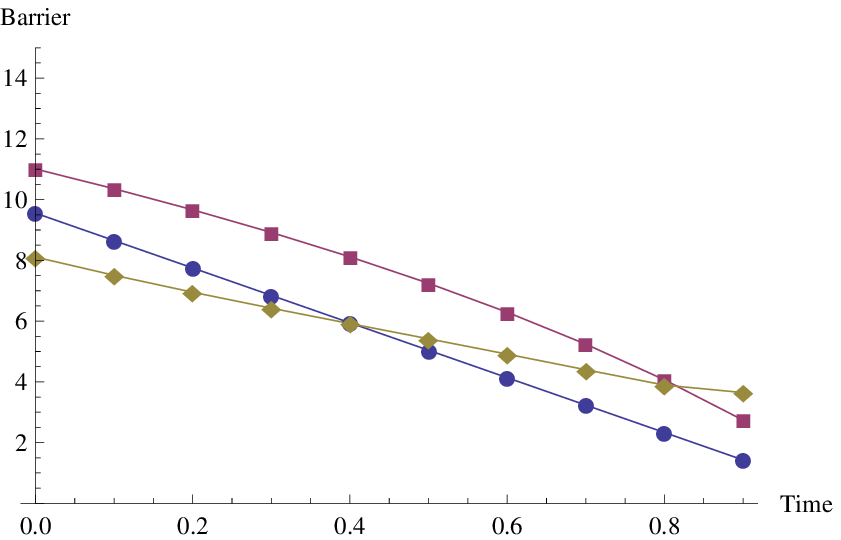} 
 \includegraphics[width=0.49\linewidth]{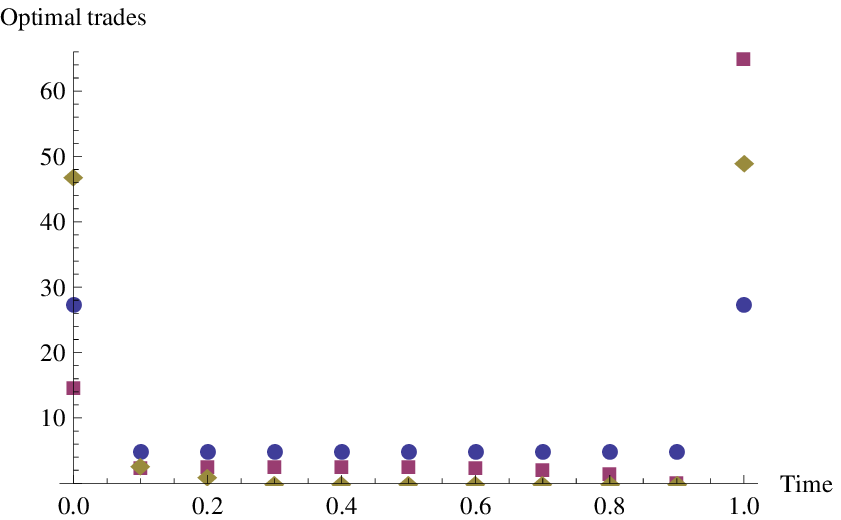} 
 \includegraphics[width=0.49\linewidth]{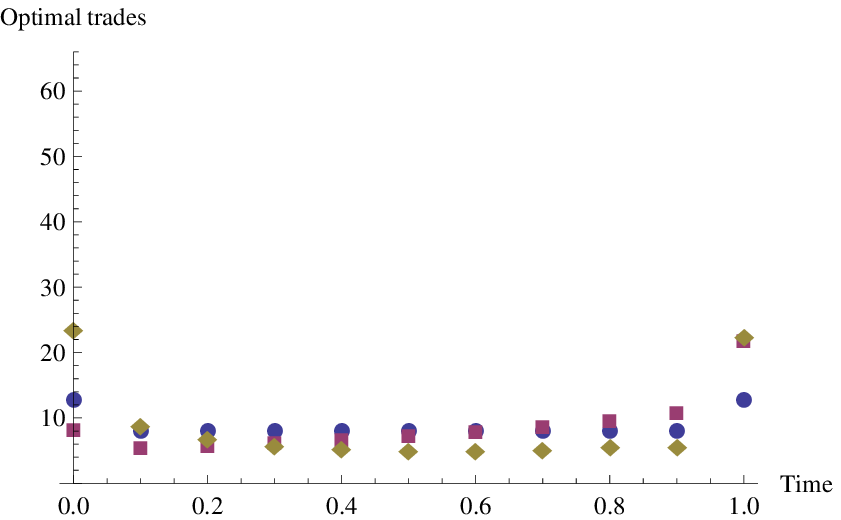}
 \caption{Illustration of the numerically computed barrier~$(c^N(t_n))_{n=0,...,N}$ and the corresponding optimal strategy~$(\De \T_{t_n})_{n=0,...,N}$ in discrete time for~$T=1$, $N=10$, $x=100$, $\de=0$, $\rho=2$ (left-hand side) and $\rho=10$ (right-hand side). We used $K^1_t \equiv 0.7$, $K^2_t = 1-0.6 t$ and $K^3_t = 1-2.4 (t-0.5)^2$ as the given evolution of the illiquidity.}
 \label{Discrete figure}
\end{figure}

We close this section with a numerical example.
Figure~\ref{Discrete figure} was generated using the above numerical scheme and illustrates the optimal barrier and trading strategy for several example definitions of~$K$ and~$\rho$. For constant~$K$, we recover the \citeasnoun{OW} ``bathtub'' strategy with impulse trades of the same size at the beginning and end of the trading horizon and trading with constant speed in between. The corresponding barrier is a decreasing straight line as we will explicitly see for continuous time in Example~\ref{example constant}. For \emph{high} values of the resilience~$\rho$, the barriers have the typical decreasing shape, i.e. the buy region increases if less time to maturity remains. For \emph{low} values of the resilience~$\rho$, the barrier must not be decreasing and can even be infinite, i.e. the buy region is the empty set, as illustated for~$K^3$ with less liquidity in the middle than in the beginning and the end of the trading horizon.

\section{Continuous time}\label{Subsec: WR-BR deterministic cts}
We now turn to the continuous time setting. In Section \ref{Subsec: ContTime ExOptStrat} we discuss existence of optimal strategies using Helly's compactness theorem and a uniqueness result using convexity of the value function. Thereafter in Section \ref{Subsec: ContTime WR-BR} we prove that the WR-BR result from Section~\ref{Subsec: WR-BR deterministic} carries over to continuous time.

\subsection{Existence of an optimal strategy}\label{Subsec: ContTime ExOptStrat}
In continuous time existence of an optimal strategy is not guaranteed in general.
For instance,  consider a constant resilience $\rho_t\equiv\rho>0$
and the price impact parameter $K$ following the Dirichlet-type function
\begin{equation}
\label{eq:Dirichlet impact}
K_t=\left\{
\begin{array}{cl}
1 & \text{for $t$ rational}\\
2 & \text{for $t$ irrational}
\end{array}
\right\}.
\end{equation}
In order to analyze model~\eqref{eq:Dirichlet impact},
let us first recall that in the model with a constant price impact $K_t\equiv\ka>0$
there exists a unique optimal strategy, which has a nontrivial absolutely continuous component
(see~\citeasnoun{OW} or Example~\ref{example constant} below for explicit formulas).
Approximating this strategy by strategies trading only at rational time points
we get that the value function in model~\eqref{eq:Dirichlet impact}
coincides with the value function for the price impact $K_t\equiv1$.
But there is no strategy in model~\eqref{eq:Dirichlet impact}
attaining this value because the nontrivial absolutely continuous component
of the unique optimal strategy for $K_t\equiv1$
will count with price impact $2$ instead of $1$ in the total costs.
Thus, there is no optimal strategy in model~\eqref{eq:Dirichlet impact}.\footnote{Let us,
however, note that the value function here has WR-BR structure
with the barrier from Example~\ref{example constant} with $\ka=1$.}

We can therefore hope to prove existence of optimal strategies
only under additional conditions on the model parameters.
In all of Section~\ref{Subsec: WR-BR deterministic cts} we will assume that $K$ is continuous;
the following theorem asserts that this is a sufficient condition for existence of an optimal strategy. 

\begin{theorem}\label{prop deterministic existence}{\rm (Continuous time: Existence).\\}
	Let~$K:[0,T]\rightarrow (0,\infty)$ be continuous. Then there exists an optimal strategy~$\T^* \in \cA_t(x)$, i.e. 
	$$J \left(t,\de,\T^* \right)=\inf_{\T \in \cA_t(x)} J \left(t,\de,\T \right).$$
\end{theorem}

In the proof we construct an optimal strategy as the limit of a sequence of (possibly suboptimal) strategies. Before we can turn to the proof itself, we need to establish that strategy convergence leads to cost convergence.
		
\begin{proposition}\label{lemma cts costs}{\rm (Costs are continuous in the strategy, $K$ continuous).\\}
	Let~$K:[0,T]\rightarrow (0,\infty)$ be continuous and let~$\bar{\T},(\T^n)$ 
	be strategies in~$\cA_t(x)$ with $\T^n \stackrel{w}{\rightarrow} 
	\bar{\T}$, i.e., $\lim_{n \rightarrow \infty} \T^{n}_s =\bar{\T}_s$ for every point~$s \in [t,T]$ of continuity of~$\bar{\T}$ (i.e.~$\T^n$  
	converges weakly to~$\bar{\T}$). Then
	$$\left \vert J(t,\de,\bar{\T})-J(t,\de,\T^n) \right \vert \xrightarrow[n \rightarrow \infty]{}0.$$
\end{proposition}

Note that Proposition~\ref{lemma cts costs} does not hold when~$K$ has a jump. To prove Proposition~\ref{lemma cts costs}, we first show in Lemma \ref{lemma cts impact} that the convergence of the price impact processes follows from the weak convergence of the corresponding strategies. We then conclude in Lemma \ref{lemma cts costs absolutely} that Proposition~\ref{lemma cts costs} holds for absolutely continuous~$K$. This finally leads to Proposition~\ref{lemma cts costs} covering all continuous~$K$.

\begin{lemma}\label{lemma cts impact}{\rm (Price impact process is continuous in the strategy).\\}
	Let~$K:[0,T]\rightarrow (0,\infty)$ be continuous and let~$\bar{\T},(\T^n)$ 
	be strategies in~$\cA_t(x)$ with $\T^n \stackrel{w}{\rightarrow} 
	\bar{\T}$.\\
	Then~$\lim_{n \rightarrow \infty} D^{n}_s =\bar{D}_s$ for~$s=T+$ and for every point~$s \in [t,T]$ of continuity of~$\bar{\T}$.
\end{lemma}

\begin{proof}
	Recall equation~(\ref{D explicit})
	$$D_s=\int_{[t,s)} K_u e^{-\int_u^s \rho_r dr} d\T_u+\de e^{-\int_t^s \rho_u du},$$
	which holds for~$s=T+$ and~$s \in [t,T]$. Due to the weak convergence (note that the total mass is preserved, i.e.~$\bar{\T}_{T+}=\T^n_{T+}=x$, 
	since~$\bar{\T},\T^n \in \cA_t(x)$) and the integrand being continuous in~$u$, the assertion follows for~$s=T+$.
	Due to the weak convergence we also have that for all~$s \in [t,T]$ with~$\De \bar{\T}_s=0$ and~$f_s(u):=K_u e^{-\int_u^s \rho_r dr} I_{[t,s)}(u)$ 
	(i.e.~$f_s$ is continuous~$d\bar{\T}$-a.e.)
	$$D^n_s= \int_{[t,T]} f_s(u)d\T^n_u  +\de e^{-\int_t^s \rho_u du} \hspace{.2cm} \xrightarrow[n \to \infty]{}
		\int_{[t,T]} f_s(u)d\bar{\T}_u +\de e^{-\int_t^s \rho_u du} =\bar{D}_s.$$
\end{proof}

\begin{lemma}\label{lemma J rewritten}{\rm (Costs rewritten in terms of the price impact process).\\}
	Let~$K:[0,T] \rightarrow (0,\infty)$ be absolutely continuous, i.e.~$K_s=K_0+\int_0^s \mu_u du$. Then
	\begin{equation}\label{D^2 form, deterministic}
		J(t,\de,\T)=\frac{1}{2} \left[\frac{D^2_{T+}}{K_T} - \frac{\de^2}{K_t} + \int_{[t,T]} \left(\frac{2 \rho_s}{K_s}+\frac{\mu_s}{K_s^2} \right) 
		D^2_s ds \right].
	\end{equation}
\end{lemma}

\begin{proof}
  Applying
	$$d\T_s=\frac{dD_s+\rho_s D_s ds}{K_s}, \: \De \T_s=\frac{\De D_s}{K_s}$$
	yields
	\begin{eqnarray*}
		J(t,\de,\T)&=&\int_{[t,T]} \left(D_s+\frac{K_s}{2} \De \T_s \right) d \T_s\\
		&=&\int_{[t,T]} \frac{D_s+\frac{1}{2}\De D_s}{K_s}dD_s+\int_{[t,T]} \frac{\rho_s D_s^2}{K_s}ds+\int_{[t,T]}\frac{\frac{1}{2}\De D_s 	
		\rho_s D_s}{K_s}ds.
	\end{eqnarray*}
	In this expression, the last term is zero since~$D$ has only countably many jumps. Using integration by parts for c\`agl\`ad processes, 
	namely~(\ref{int by parts 2}) with~$U:=D, \, V:=\frac{D}{K}$, and~$d\left(\frac{D_s}{K_s}\right)=\frac{1}{K_s}dD_s+D_s d \left(\frac{1}{K_s} 
	\right)$, we can write
	$$\int_{[t,T]} \frac{D_s}{K_s}dD_s=\frac{1}{2}\left[ \frac{D_{T+}^2}{K_T}-\frac{\de^2}{K_t}-\int_{[t,T]} D_s^2 d\left(\frac{1}{K_s}\right)-\sum_{s 
	\in [t,T]}\frac{(\De D_s)^2}{K_s}\right].$$
	Plugging in~$d \left(\frac{1}{K_s}\right)= -\frac{\mu_s}{K_s^2}ds$ yields~(\ref{D^2 form, deterministic}) as desired.
\end{proof}

The following result is a direct consequence of Lemma~\ref{lemma cts impact} and Lemma~\ref{lemma J rewritten}.
\begin{lemma}\label{lemma cts costs absolutely}{\rm (Costs are continuous in the strategy, $K$ absolutely continuous).\\}
	Let~$K:[0,T] \rightarrow (0,\infty)$ be absolutely continuous and~$\bar{\T},(\T^n)$ 
	be strategies in~$\cA_t(x)$ with~$\T^n 
	\stackrel{w}{\rightarrow} \bar{\T}$. Then
	$$\left \vert J(t,\de,\bar{\T})-J(t,\de,\T^n) \right \vert \xrightarrow[n \rightarrow \infty]{}0.$$
\end{lemma}

\begin{proof}[Proof of Proposition~\ref{lemma cts costs}]
	We use a proof by contradiction and suppose there exists a subsequence $\left(n_j \right) \subset \N$ such that
	$$\lim_{j \rightarrow \infty} \int_{[t,T]} \left(D^{n_j}_s+\frac{K_s}{2} \De \T^{n_j}_s \right) d \T^{n_j}_s
	\neq \int_{[t,T]} \left(\bar{D}_s+\frac{K_s}{2} \De \bar{\T}_s \right) d \bar{\T}_s,$$
	where the limit on the left-hand side exists. Without loss of generality assume
	\begin{equation}\label{J contradiction}
		\lim_{j \rightarrow \infty} \int_{[t,T]} \left(D^{n_j}_s+\frac{K_s}{2} \De \T^{n_j}_s \right) d \T^{n_j}_s
	< \int_{[t,T]} \left(\bar{D}_s+\frac{K_s}{2} \De \bar{\T}_s \right) d \bar{\T}_s.
	\end{equation}
	We now want to bring Lemma~\ref{lemma cts costs absolutely} into play. For~$\ep >0$, we denote by~$K^{\ep}:[t,T] \rightarrow (0,\infty)$ an 
	absolutely continuous function such that~$\max_{s \in [t,T]} \left|K^{\ep}_s - K_s \right| \leq \ep$. For~$\T \in \cA_t(x)$
	\begin{eqnarray*}
		&& \left|\int_{[t,T]} \left(D^{\ep}_s+\frac{K^{\ep}_s}{2} \De \T_s \right) d \T_s -
					   \int_{[t,T]} \left(D_s+\frac{K_s}{2} \De \T_s \right) d \T_s \right| \\
		&\leq&  \int_{[t,T]} \left(\left|D^{\ep}_s-D_s \right| + \frac{1}{2} \left|K^{\ep}_s-K_s \right| \De \T_s \right) d \T_s \leq \frac{3}{2} x^2 \ep.
	\end{eqnarray*}
	We therefore get from~(\ref{J contradiction}) that there exists~$\ep >0$ such that
	$$\limsup_{j \rightarrow \infty} \int_{[t,T]} \left(D^{n_j,\ep}_s+\frac{K^{\ep}_s}{2} \De \T^{n_j}_s \right) d \T^{n_j}_s
	< \int_{[t,T]} \left(\bar{D}^{\ep}_s+\frac{K^{\ep}_s}{2} \De \bar{\T}_s \right) d \bar{\T}_s.$$
	This is a contradiction to Lemma~\ref{lemma cts costs absolutely}.
\end{proof}

We can now conclude the proof of the existence Theorem \ref{prop deterministic existence}.

\begin{proof}[Proof of Theorem \ref{prop deterministic existence}]
	Let~$(\T^n) \subset \cA_t(x)$ be a minimizing sequence. Due to the monotonicity of the considered strategies, we can use 
	\emph{Helly's Theorem} in the form of Theorem~2, \S 2, Chapter III of \citeasnoun{Shiryaev}, which also holds for left-continuous processes and 
	on~$[t,T]$ instead of~$(-\infty,\infty)$. It guarantees the existence of a deterministic~$\bar{\T} \in	\cA_t(x)$ 
	and a subsequence~$(n_j) 
	\subset \N$ such that~$(\T^{n_j})$ converges weakly to~$\bar{\T}$. Note that we can always force~$\bar{\T}_{T+}$ to be~$x$, since weak 
	convergence does not require that~$\T^{n_j}_T$ converges to~$\bar{\T}_T$ whenever~$\bar{\T}$ has a jump at~$T$. Thanks to Proposition~\ref{lemma 
	cts costs}, we can conclude that
	$$U(t,\de,x)=\lim_{j \rightarrow \infty} J(t,\de,\T^{n_j})=J(t,\de,\bar{\T}).$$
\end{proof}

The price impact process~$D$ is affine in the corresponding strategy~$\T$. That is in the case when $K$ is not decreasing too quickly, Lemma~\ref{lemma J rewritten} guarantees that the cost term~$J$ is strictly convex in the strategy~$\T$. Therefore, we get the following uniqueness result.

\begin{theorem}\label{thm deterministic uniqueness}{\rm (Continuous time: Uniqueness).\\}
	Let~$K:[0,T] \rightarrow (0,\infty)$ be absolutely continuous, i.e.~$K_s=K_0+\int_0^s \mu_u du$, and additionally~
	$$\mu_s+2 \rho_s K_s>0 \, \text{ a.e. on } \,[0,T] \, \text{ with respect to the Lebesgue measure.}$$
	Then there exists a unique optimal strategy.
\end{theorem}
\subsection{WR-BR structure}\label{Subsec: ContTime WR-BR}
For continuous~$K$, we have now established existence and (under additional conditions) uniqueness of the optimal strategy. Let us now turn to the value function in continuous time and demonstrate that it has WR-BR structure, consistent with our findings in discrete time. 

\begin{theorem}\label{prop deterministic WR-BR cts}{\rm (Continuous time: WR-BR structure).\\}
	Let~$K:[0,T]\rightarrow (0,\infty)$ be continuous. Then the value function has WR-BR structure. 
\end{theorem}
 
We are going to deduce the structural result for the continuous time setting by using our discrete time result. First, we show that the discrete time value function converges to the continuous time value function. Without loss of generality, we set~$t=0$.

\begin{lemma}\label{lemma deterministic cts WRBR 1}
{\rm (The discrete time value function converges to the continuous time one).\\}
	Let~$K:[0,T]\rightarrow (0,\infty)$ be continuous and consider an equidistant time grid with~$N$ trading intervals. Then
	$$\lim_{N \rightarrow \infty} V^N(0,y)=V(0,y).$$
\end{lemma}

\begin{proof}
	Thanks to Theorem~\ref{prop deterministic existence}, there exists a continuous time optimal strategy~$\T^* \in \cA_0(y)$. Appro\-xi\-mate it 
	suitably via step functions~$\T^N \in \cA_0^N(y)$. 
	Then
	\begin{eqnarray*}
		V(0,y) &=& J(0,1,\T^*)=\lim_{N \rightarrow \infty} J(0,1,\T^N) \geq  \limsup_{N \rightarrow \infty} V^N(0,y).
	\end{eqnarray*}
	The inequality~$V(0,y) \leq \liminf_{N \rightarrow \infty}  V^N(0,y)$ is immediate.
\end{proof}

\begin{proof}[Proof of Theorem~\ref{prop deterministic WR-BR cts}]
	By the same change of variable from~$\xi$ to~$\eta$ that was used in Section \ref{Subsec: WR-BR deterministic}, we can transform the optimal trade equation 
$$ V(0,y) = \min_{\xi \in [0,y]} \left\{ \left(1+\frac{K_0}{2} \xi \right) \xi+(1+ K_0 \xi)^2 V\left(0,\frac{y-\xi}{1+K_0 \xi} \right)\right\}$$
into the optimal barrier equation
\begin{equation}\label{equation V1}
 V(0,y) =  \frac{1}{2 K_0} \left[(1+K_0 y)^2 \min_{\eta \in [0,y]} L(0,\eta)-1 \right],
\end{equation}
	where
	\begin{equation}\label{L in cts time}
		L(0,y):=L(y):=\frac{1+2 K_0 V(0,y)}{(1+K_0 y)^2}.
	\end{equation}
	Now it follows from~(\ref{equation V1}) and~(\ref{L in cts time}) that
	$$\min_{\eta \in [0,y]} L(\eta)=L(y),$$
	in particular the function~$L$ is nonincreasing in~$y$. 
Define
$$\tilde{L}^N(y):=\min_{\eta \in [0,y]} L^N(0,\eta),$$
which is a nonincreasing positive function.
If for some~$N$ the function $y\mapsto L^N(0,y)$ is not convex on $[0,c^N(0))$,
then the second alternative in Lemma~\ref{lemma deterministic cts WRBR 2} holds,
i.e. we have WR-BR structure with $c(0)=\infty$.
Thus, below we assume that for any~$N$ the function $y\mapsto L^N(0,y)$ is convex on $[0,c^N(0))$,
hence, by Lemma~\ref{lemma L}~(a) and Theorem~\ref{prop deterministic WR-BR},
the function $\tilde L^N$ is convex on $[0,\infty)$.
Moreover, by rearranging~(\ref{optimal barrier eq}) we obtain that
	$$\tilde{L}^N(y)=\frac{1+2K_0 V^N(0,y)}{(1+K_0 y)^2}.$$
	Hence~$\tilde{L}^N$ converges pointwise to~$L$ as~$N \rightarrow \infty$ by Lemma~\ref{lemma deterministic cts WRBR 1} and~(\ref{L in cts time}). 
	Therefore,~$L$ is also convex. 
	
Due to~$L$ being nonincreasing and convex,
there	exists a unique~$c^* \in [0,\infty]$ such that~$L$ is strictly decreasing 
for~$y \in [0,c^*)$ and constant for~$y \in (c^*,\infty)$. One can now conclude that for all~$y>c^*$ and~$\eta \in \left(c^*,y\right)$, 
	setting~$\xi:=\frac{y-\eta}{1+K_0 \eta}$, i.e.~$\eta=\frac{y-\xi}{1+K_0 \xi}$, and using~(\ref{L in cts time}) and~$L(y)=L(\eta)$, we have
	\begin{eqnarray*}
		V(0,y)&=&\frac{1}{2 K_0} \left[(1+K_0 y)^2 L(y)-1 \right]\\
		&=&\frac{1}{2 K_0} \left[(1+K_0 y)^2 L(\eta)-1 \right]\\
		&=&\frac{1}{2 K_0} \left[(1+K_0 y)^2 L\left(\frac{y-\xi}{1+K_0 \xi}\right)-1 \right].
	\end{eqnarray*}
	We now use the definition of~$L$ from~(\ref{L in cts time}) once again to get
	\begin{equation}\label{V eq 355}
		V(0,y)=\left(1+\frac{K_0}{2} \xi \right) \xi+(1+ K_0 \xi)^2 V\left(0,\frac{y-\xi}{1+K_0 \xi} \right).
	\end{equation}
	Therefore $\left(c^*,\infty \right)\subset Br_0$. In case of~$c^*>0$, consider~$y \leq c^*$, take any~$\eta \in [0,y)$, and set~$\xi:=\frac{y-\eta}{1+K_0 \eta}$. 
	Then a similar calculation using that~$L(y)<L(\eta)$ shows that~$V(0,y)$ is strictly smaller than the right-hand side of~(\ref{V eq 355}). 
	Hence~$Br_0=\left(c^*,\infty \right)$.
	Thus, we get WR-BR structure with $c(0)=c^* \in [0,\infty]$ as desired.
\end{proof}

In Section~\ref{Subsec: Euler Lagrange} we will investigate the value function, barrier function and optimal trading strategies for several example specifications of~$K$ and~$\rho$.
\section{Zero spread and price manipulation}\label{SecZeroSpreadPriceManipulation}

In the model introduced in Section~\ref{SecModel}, we assumed a trading dependent spread between the best ask~$A_t$ and best bid~$B_t$. This has allowed us to exclude both forms of price manipulation in Section~\ref{SecMarketManipulation}. An alternative assumption that is often made in limit order book models is to disregard the bid-ask spread and to assume~$A_t = B_t$, see, for example, \citeasnoun{Huberman2004}, \citeasnoun{Gatheral}, \citeasnoun{ASS} and \citeasnoun{GatheralSchiedSlynko1}. The canonical extension of these models to our framework including time-varying liquidity is the following.

\begin{assumption}\label{Assum zero spread}
In the \emph{zero spread model}, we have the \emph{unaffected price}~$S^u$, which is a c\`adl\`ag $\cH^1$-martingale with a deterministic starting point~$S^u_0$, and assume that the best bid and ask are equal and given by~$A_t^\updownarrow = B_t^\updownarrow = S_t^u + D_t^\updownarrow$ with
\begin{equation}\label{D zero spread}
D_t^\updownarrow=D^\updownarrow_0e^{-\int_0^t\rho_s\,ds}+\int_{[0,t)}K_se^{-\int_s^t\rho_u\,du}(d\T_s-d\tT_s),\quad t\in[0,T+],
\end{equation}
where $D^\updownarrow_0\in\R$ is the initial value for the price impact. For convenience, we will furthermore assume that $K\colon[0,T]\to(0,\infty)$ is twice continuously differentiable and $\rho\colon[0,T]\to(0,\infty)$ is continuously differentiable.
\end{assumption}

As opposed to this zero spread model, the model introduced in Section~\ref{SecModel} will be referred to as the \emph{dynamic spread model} in the sequel. In this section we study price manipulation and optimal execution in the zero spread model. In particular, we provide explicit formulas for optimal strategies. This in turn will be used in the next section to study explicitly several examples both in the dynamic spread and in the zero spread model.

We have excluded permanent impact from the definition above ($\ga=0$). It can easily be included, but, like in the dynamic spread model, proves to be irrelevant for optimal strategies and price manipulation. Note that for pure buying strategies ($\tT\equiv0$) the zero spread model is identical to the model introduced in Section~\ref{SecModel}. The difference between the two models is that if sell orders occur, then they are executed at the same price as the ask price. Furthermore, buy and sell orders impact this price symmetrically. We can hence consider the net trading strategy $\Tud:=\T-\tT$ instead of buy orders~$\T$ and sell orders~$\tT$ separately. The simplification of the stochastic optimization problem of Section~\ref{Subsec: Trader objectives} to a deterministic problem in Section~\ref{sec:red_opt_prob} applies similarly to the zero spread model defined in Assumption~\ref{Assum zero spread}.
Thus, for any fixed $x\in\R$, we define the sets of strategies
\begin{align*}
\cA^\updownarrow&:=\big\{\Tud\colon[0,T+]\rightarrow\R\,|\,\Tud
\text{ is a deterministic c\`agl\`ad}\\
&\hspace{9mm}\text{function of finite variation with }\Tud_0=0\big\},\\
\cA^\updownarrow(x)&:=\left\{\Tud\in\cA^\updownarrow\,|\,\Tud_{T+}=x\right\}.
\end{align*}
Strategies from $\cA^\updownarrow(x)$ allow buying and selling and build up the position of $x$ shares until time~$T$.
We further define the cost function~$J^\updownarrow\colon\R\times\cA^\updownarrow\rightarrow\R$ as
$$
J^\updownarrow(\Tud):=J(\de,\Tud):= \int_{[0,T]} \left(D^\updownarrow_s+\frac{K_s}{2} \De \Tud_s \right) d\Tud_s,
$$
where $D^\updownarrow$ is given by~\eqref{D zero spread} with $D^\updownarrow_0=\de$.
The function~$J^\updownarrow$ represents the total temporary impact costs\footnote{In the case of liquidation of shares
(i.e.~$\Tud_{T+}<0$) the word ``costs'' should be understood as ``minus proceeds from the liquidation''.}
in the zero spread model of the strategy~$\Tud$
on the time interval~$[0,T]$ when the initial price impact $D^\updownarrow_0=\de$.
Observe that $J^\updownarrow$ is well-defined and finite because $K$ is bounded,
which in turn follows from Assumption~\ref{Assum zero spread}.
The value function $U^\updownarrow\colon\R^2\rightarrow\R$ is then given as
\begin{equation}
\label{eq:zspm value fun}
U^\updownarrow(\de,x):=\inf_{\Tud \in \cA^\updownarrow(x)} J^\updownarrow(\de,\Tud).
\end{equation}

The zero spread model admits \emph{price manipulation} if, for~$D^\updownarrow_0=0$,
there is a profitable round trip, i.e. there is $\Tud\in\cA^\updownarrow(0)$ with $J^\updownarrow(0,\Tud)<0$.
The zero spread model admits \emph{transaction-triggered price manipulation} if, for~$D^\updownarrow_0=0$,
the execution costs of a buy (or~sell) program can be decreased by intermediate sell (resp.~buy) trades
(more precisely, this should be formulated like in Definition~\ref{def:ttpm1}).

\begin{remark}
\label{rem:zspm1}
The conceptual difference with Section~\ref{SecMarketManipulation} is that we require
$D^\updownarrow_0=0$ in these definitions.
The reason is that even in ``sensible'' zero spread models
(that do not admit both types of price manipulation according to definitions above),
we typically have profitable round trips whenever $D^\updownarrow_0\ne0$.
In the zero spread model, the case $D^\updownarrow_0\ne0$
can be interpreted as that the market price is not in its equilibrium state
in the beginning. In the absence of trading the process $(D^\updownarrow_t)$
approaches zero due to the resilience, hence both best ask and best bid price processes
$(A^\updownarrow_t)$ and $(B^\updownarrow_t)$ (which are equal)
approach their evolution in the equilibrium~$(S^u_t)$.
The knowledge of this ``direction of deviation from~$S^u$''
plus the fact that both buy and sell orders are executed at the same price\footnote{Let us observe
that this does not apply to the dynamic spread model of Section~\ref{SecModel},
where we have different processes $D$ and $E$ for the deviations of the best ask and best bid prices
from the unaffected ones due to the previous trades.}
clearly allow us to construct profitable round trips.
For instance, in the Obizhaeva--Wang-type model with a constant price impact $K_t\equiv\ka>0$,
the strategy
$$
\Tud_s:=-\frac{D^\updownarrow_0}{2\ka}I_{(0,\ep]}(s),\quad s\in[0,T+],
$$
where $\ep\in(0,T]$, is a profitable round trip whenever $D^\updownarrow_0\ne0$,
as can be checked by a straightforward calculation.
\end{remark}

Let us first discuss classical price manipulation in the zero spread model.
If the liquidity in the order book rises too fast ($K$ falls too quickly), then a simple pump and dump strategy becomes attractive.
In the initial low liquidity regime (high~$K$), buying a large amount of shares increases the price significantly.
Quickly thereafter liquidity increases.
Then the position can be liquidated with little impact at this elevated price, leaving the trader with a profit.
The following result formalizes this line of thought.

\begin{proposition}\label{Thm zero spread pm}{\rm (Price manipulation in the zero spread model).\\}
Assume the zero spread model of Assumption~\ref{Assum zero spread} and that
$$
K'_t+2\rho_t K_t<0\quad\text{for some }t\in[0,T).
$$
Then price manipulation occurs and, for any $\de,x\in\R$,
there is no optimal strategy in problem~\eqref{eq:zspm value fun}.
\end{proposition}

\begin{proof}
By the assumption of the theorem,
$$
K'_t=\lim_{\ep\searrow0}\frac{K_{t+\ep}-K_t}\ep<-2\rho_tK_t
=\lim_{\ep\searrow0}\frac{K_t\left(2e^{-\int_t^{t+\ep}\rho_u\,du}-1\right)-K_t}\ep,
$$
hence for a sufficiently small $\ep>0$ we have
\begin{equation}
\label{eq:pmzs1}
K_{t+\ep}<K_t\left(2e^{-\int_t^{t+\ep}\rho_u\,du}-1\right).
\end{equation}
Let us consider the round trip $\T^{\updownarrow m}\in\cA^\updownarrow(0)$,
which buys $1$ share at time~$t$ and sells it at time~$t+\ep$, i.e.
$$
\T^{\updownarrow m}_s:=I_{(t,t+\ep]}(s),\quad s\in[0,T+].
$$
A straightforward computation shows that, for~$D^\updownarrow_0=0$, the cost of such a round trip is
$$
J^\updownarrow(0,\T^{\updownarrow m})
=\frac{K_t+K_{t+\ep}}2-K_te^{-\int_t^{t+\ep}\rho_u\,du}.
$$
Due to~\eqref{eq:pmzs1}, $J^\updownarrow(0,\T^{\updownarrow m})<0$. Thus, price manipulation occurs.

Let us fix $\de,x\in\R$, consider a strategy $\Tud\in\cA^\updownarrow(x)$, and, for any $z\in\R$, define
$$
\T^{\updownarrow z}:=\Tud+z\T^{\updownarrow m}.
$$
Then $\T^{\updownarrow z}\in\cA^\updownarrow(x)$ and we have
$$
J^\updownarrow(\de,\T^{\updownarrow z})=c_0z^2+c_1z+c_2
$$
with $c_0=J^\updownarrow(0,\T^{\updownarrow m})<0$ and some constants $c_1$ and~$c_2$.
Since $z$ is arbitrary, we get $U^\updownarrow(\de,x)=-\infty$.
An optimal strategy is this situation would be a strategy from $\cA^\updownarrow(x)$ with the cost~$-\infty$.
But for any strategy~$\Tud$, its cost $J^\updownarrow(\de,\Tud)$ is finite as discussed above,
hence there is no optimal strategy in problem~\eqref{eq:zspm value fun}.
\end{proof}

Interestingly, the condition $K'_t+2\rho_t K_t<0$ for some~$t$ in Proposition~\ref{Thm zero spread pm} is not symmetric; quickly falling~$K$ leads to price manipulation, but quickly rising~$K$ does not. 

If~$K'_t+2\rho_t K_t\geq 0$ holds at all points in time, then the situation remains unclear so far.
In their model, \citeasnoun{ASS} and \citeasnoun{GatheralSchiedSlynko1} have shown that even in the absence of profitable round trip trades, we might still be facing transaction-triggered price manipulation.
This can happen also in our zero spread model.
The following theorem provides explicit formulas for optimal strategies and leads to a characterization of transaction-triggered price manipulation.

\begin{theorem}\label{Thm zero spread opt strat}{\rm (Optimal strategies in the zero spread model).\\}
Assume the zero spread model of Assumption~\ref{Assum zero spread} and that $K'_t+2 \rho_t K_t > 0$ on~$[0,T]$. Define
\begin{equation}
\label{eq:zspm f eq}
f_t:=\frac{K'_t+\rho_tK_t}{K'_t+2\rho_t K_t},\quad t\in[0,T].
\end{equation}
Then, for any $\delta,x\in\R$, the strategy $\T^{\updownarrow*}$ given by the formulas
\begin{equation}\label{closed form optimal strategy}
\De \T^{\updownarrow *}_0=\de^\updownarrow \frac{f_0}{K_0}-\frac{\de}{K_0}, \hspace{.5 cm} 
d\T^{\updownarrow *}_t=\de^\updownarrow \frac{f'_t+\rho_t f_t}{K_t} dt, \hspace{.5 cm} 
\De	\T^{\updownarrow *}_T=\de^\updownarrow\frac{1-f_T}{K_T}
\end{equation}
with $\de^\updownarrow:=\frac1c\left(x+\frac\de{K_0}\right)$ and
\begin{equation}
\label{eq:zspm c greater 0}
c:=\int_0^T\frac{f'_t+\rho_t f_t}{K_t} dt+\frac{f_0}{K_0}+\frac{1-f_T}{K_T} >0,
\end{equation}
is the unique optimal strategy in problem~\eqref{eq:zspm value fun}.
Furthermore, we have
\begin{equation}
\label{eq:zspm opt value}
U^\updownarrow (\delta,x)=J^\updownarrow(\de,\T^{\updownarrow*})
=(\de^\updownarrow)^2\left(\int_0^T(K'_t+2\rho_tK_t)\frac{f_t^2}{2K_t^2}\,dt+\frac1{2K_T}\right)-\frac{\de^2}{2K_0}.
\end{equation}
\end{theorem}

\begin{corollary}[Transaction-triggered price manipulation in the zero spread model]
\label{cor:zspm ttpm}\mbox{}\\
Under the assumptions of Theorem~\ref{Thm zero spread opt strat} price manipulation does not occur.
Furthermore, transaction-triggered price manipulation occurs if and only if $f_0<0$ or $f'_t+\rho_t f_t<0$ for some~$t\in[0,T]$.
\end{corollary}

\begin{proof}
Using~\eqref{closed form optimal strategy} with~$\de=0$, we immediately get that price manipulation does not occur.
Noting further that~$f_T<1$, we obtain that transaction-triggered price manipulation occurs if and only if either~$f_0<0$ or~$f'_t+\rho_tf_t<0$ for some~$t\in[0,T]$.
\end{proof}

\begin{figure}[htbp]
\centering
\includegraphics[width=0.49\linewidth]{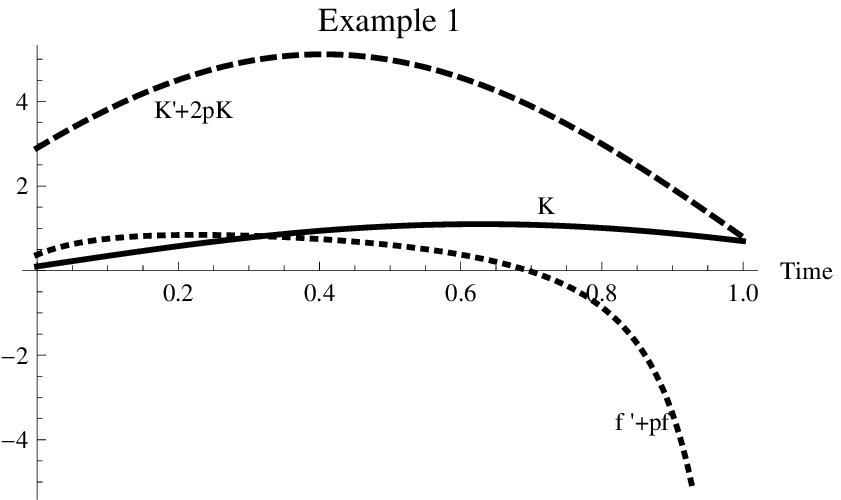}
\includegraphics[width=0.49\linewidth]{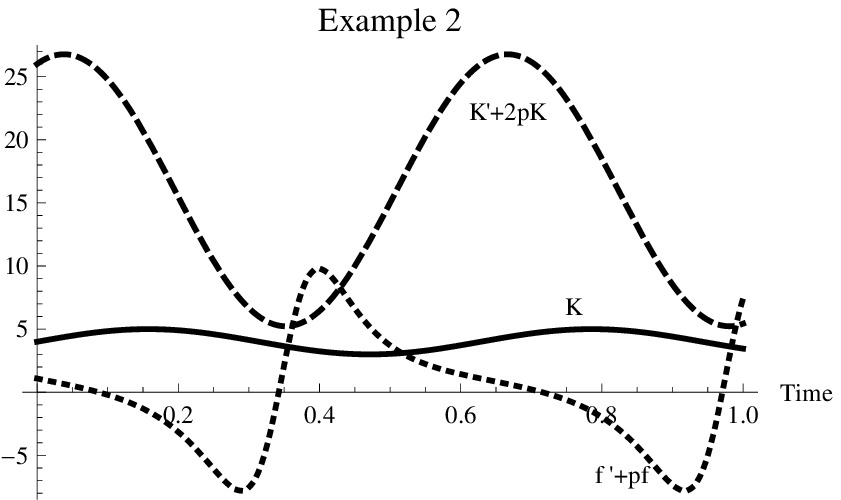}
\includegraphics[width=0.49\linewidth]{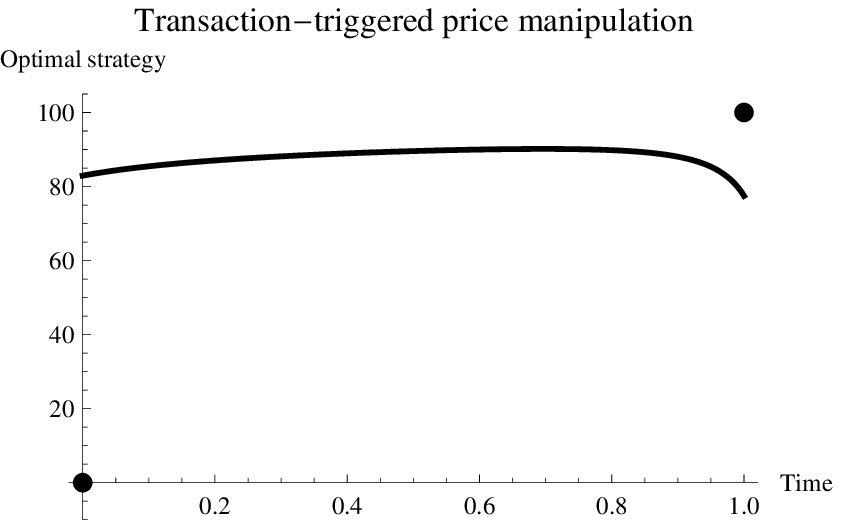}
\includegraphics[width=0.49\linewidth]{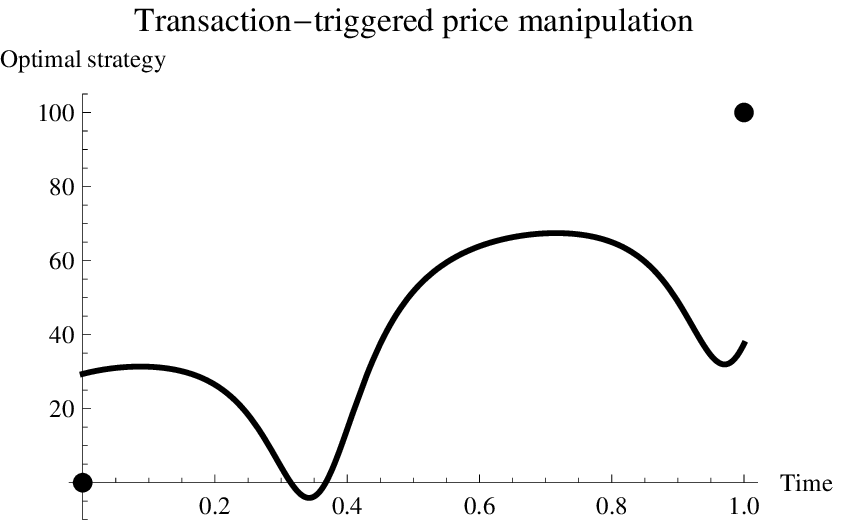}
\caption{In Example~1, we consider~$K_t=\sin(2.5 t)+0.1$ and~$K_t=\sin(10 t)+4$ in Example~2. The other parameters are~$T=1$, $\rho=2$, $x=100$, $\de=0$. The plots at the bottom illustrate the corresponding optimal
strategies~$\T^{\updownarrow *}$ from~\eqref{closed form optimal strategy}.
}
\label{transaction-triggered figure}
\end{figure}

We can summarize Proposition~\ref{Thm zero spread pm} and Corollary~\ref{cor:zspm ttpm} as follows.
If $K'_t+2\rho_t K_t<0$ for some~$t$, i.e. liquidity grows very rapidly, then price manipulation (and hence transaction-triggered price manipulation) occurs.
If $K'_t+2\rho_t K_t>0$ everywhere, but $f_0<0$ or $f'_t+\rho_t f_t<0$ for some~$t$,
i.e. liquidity grows fast but not quite as fast,
then price manipulation does not occur, but transaction-triggered price manipulation occurs.
If $K'_t+2\rho_t K_t>0$ everywhere, $f_0\ge0$
and $f'_t+\rho_t f_t\ge0$ everywhere,
i.e. liquidity never grows too fast, then neither form of price manipulation occurs and an investor wishing to purchase should only submit buy orders to the market.
Figure~\ref{transaction-triggered figure} illustrates optimal transaction-triggered price manipulation strategies.
In Example~1, the liquidity~$\frac{1}{K}$ is slightly growing at the end of the trading horizon, which makes the optimal
strategy~$\T^{\updownarrow *}$
non-monotonic. As we see in Example~2, the number of shares hold by the large investor can become negative although the overall goal is to buy a positive amount of shares.

In the proof of Theorem~\ref{Thm zero spread opt strat}, we are going to exploit the fact that there is a one-to-one correspondence between $\Tud$ and~$D^\updownarrow$. We rewrite the cost term, which is essentially $\int_0^T D^\updownarrow_t d\Tud_t$, in terms of the deviation process~$D^\updownarrow$ by applying
\begin{equation}\label{Theta via D}
d\Tud_t=\frac{dD^\updownarrow_t+\rho_t D^\updownarrow_t dt}{K_t},
\end{equation}
and get the following result.

\begin{lemma}[Costs rewritten in terms of the price impact process]
\label{lem:zspm costs}\mbox{}\\
Under Assumption~\ref{Assum zero spread}, for any $\de\in\R$ and $\Tud\in\cA^\updownarrow$, we have
\begin{equation}\label{eq:zspm costs}
J^\updownarrow(\de,\Tud)=\frac12\left[\frac{\left(D^\updownarrow_{T+}\right)^2}{K_T}-\frac{\de^2}{K_0}
+\int_{[0,T]}\left(K'_t+2\rho_tK_t\right)\frac{\left(D^\updownarrow_t\right)^2}{K_t^2}\,dt\right].
\end{equation}
\end{lemma}

The formal proof, where one needs to take into account possible jumps of~$\Tud$,
is similar to that of Lemma~\ref{lemma J rewritten}.

Similar to \citeasnoun{BB} and as explained in \citeasnoun{GregoryLin}, we can then use the Euler-Lagrange formalism to find necessary conditions on the optimal~$D^\updownarrow$. Under our assumptions, these conditions turn out to be sufficient and the optimal~$D^\updownarrow$ directly gives us an optimal~$\Tud$. Unfortunately, we cannot use the Euler-Lagrange approach directly in the full generality of all strategies in~$\cA^\updownarrow$, but need to impose a continuity property. Motivated by the WR-BR structure established in previous sections as well as the optimal strategy in the case of constant~$K$ from~\citeasnoun{OW}, we introduce, for~$x\in\R$, the set of strategies~$\cA^{\updownarrow	c}(x)\subset\cA^\updownarrow(x)$ with impulse trades at~$t=0$ and~$t=T$ only:
$$
\cA^{\updownarrow	c}(x):=\big\{\Tud\in\cA^\updownarrow(x)\,|\,\Tud\text{ is continuous on~$(0,T)$}\big\}.
$$
We will also need a notation for a similar set of monotonic strategies, i.e., for~$y\in[0,\infty)$, we define
$$
\cA^c_0(y):=\big\{\T\in\cA_0(y)\,|\,\T\text{ is continuous on~$(0,T)$}\big\}.
$$

\begin{lemma}\label{cts strategies cor}{\rm (Approximation by continuous strategies).\\}
Assume the zero spread model of Assumption~\ref{Assum zero spread}. Then, for any $\de,x\in\R$,
\begin{equation}
\label{eq:zspm infs equal}
U^\updownarrow(\de,x) := \inf_{\Tud\in\cA^\updownarrow(x)}J^\updownarrow(\de,\Tud)
=\inf_{\Tud\in\cA^{\updownarrow c}(x)}J^\updownarrow(\de,\Tud).
\end{equation}
\end{lemma}

\begin{proof}
Let us take any $\Tud\in\cA^\updownarrow(x)$
and find $\T,\tT\in\cA_0$ such that $\Tud=\T-\tT$.
We set $y:=\T_{T+}\in[0,\infty)$, $\tilde y:=\tT_{T+}\in[0,\infty)$,
so that $x=y-\tilde y$.
Below we will show that
\begin{equation}
\label{eq:zspm theta_n}
\exists\,\T^n\in\cA^c_0(y),\;\tT^n\in\cA^c_0(\tilde y)\text{ such that }
\T^n\stackrel{w}{\longrightarrow}\T,\;\tT^n\stackrel{w}{\longrightarrow}\tT.
\end{equation}
Let us define $\T^{\updownarrow n}:=\T^n-\tT^n\in\cA^{\updownarrow c}(x)$.
It follows from~\eqref{D zero spread} and the weak convergence of the strategies
that the price impact~$D^{\updownarrow n}_t$ corresponding to~$\T^{\updownarrow n}$
converges to the price impact~$D^\updownarrow_t$ corresponding to~$\Tud$
for $t=T+$ and for every point $t\in[0,T]$, where both $\T$ and $\tT$ are continuous
(i.e.~the convergence of price impact functions holds at~$T+$
and everywhere on $[0,T]$ except at most a countable set).
By~\eqref{eq:zspm costs}, we get
$J^\updownarrow(\de,\T^{\updownarrow n})\to J^\updownarrow(\de,\Tud)$ as~$n \rightarrow \infty$.
Since $\Tud\in\cA^\updownarrow(x)$ was arbitrary, we obtain~\eqref{eq:zspm infs equal}.

It remains to prove~\eqref{eq:zspm theta_n}.
Clearly, it is enough to consider some $\T\in\cA_0(y)$ and to construct
$\T^n\in\cA^c_0(y)$ weakly convergent to~$\T$.
Let~$\cP$ denote the class of all probability measures~$P$ on~$([0,T],\cB([0,T]))$ and
$$\cP^c= \left\{P \in \cP \big| P(\{s\})=0 \text{ for all } s \in (0,T) \right\}.$$
The formula $P\left([0,s) \right):=\frac{\T_s}{y}$, $s \in [0,T]$,
with~$\T \in \cA_0(y)$, provides a one-to-one correspondence between~$\cA_0(y)$ and~$\cP$, where~$\cA^c_0(y)$ is mapped on~$\cP^c$.
Thus, it is enough to show that any probability measure~$P \in \cP$ can be weakly approximated by probability measures from~$\cP^c$.
To this end, let us consider independent random variables~$\psi$ and~$\zeta$ such that~$\text{Law}(\psi)=P$ and~$\text{Law}(\zeta)$ is continuous. For any~$n \in \N$, we define
$$\psi_n:=\left(\left(\psi+\frac{\zeta}{n}\right)\vee 0\right) \wedge T.$$
Then
$$Q_n:=\text{Law}(\psi_n) \in \cP^c$$
and~$Q_n \stackrel{w}{\rightarrow} P$ as~$n \rightarrow \infty$ because~$\psi_n \rightarrow \psi$ a.s.
This concludes the proof.
\end{proof}

\begin{lemma}\label{lemma c>0}
	Assume~$K'_t + 2\rho_t K_t > 0$ on~$[0,T]$ and define
	$$\chi(t) := \int_0^t\frac{f'_s+\rho_s f_s}{K_s}dt + \frac{f_0}{K_0} + \frac{1-f_t}{K_t}.$$
	Then~$\chi(t) > 0$ for all~$t\in [0,T]$. In particular,~$c=\chi(T)>0$.
\end{lemma}

\begin{proof}
	We have 
	$$\chi(0) = \frac{1}{K_0} > 0.$$
	Furthermore,
	$$\chi'(t) = \frac{f'_t+\rho_t f_t}{K_t} + \frac{-f'_t K_t - (1-f_t)K'_t}{K_t^2} = \frac{\rho_t^2}{K'_t + 2\rho_t K_t} >0.$$
\end{proof}

\begin{proof}[Proof of Theorem~\ref{Thm zero spread opt strat}]
We first note that~$c$ from~(\ref{eq:zspm c greater 0}) is strictly positive by Lemma~\ref{lemma c>0}. Also note that if an optimal strategy in~\eqref{eq:zspm value fun} exists, then it is unique in the class~$\cA^\updownarrow(x)$ because the function~$\Tud\mapsto J^\updownarrow(\de,\Tud)$
is strictly convex on~$\cA^\updownarrow$ (this is due to~\eqref{eq:zspm costs} and the assumption~$K'_t+2\rho_tK_t>0$ on~$[0,T]$).

For the strategy~$\T^{\updownarrow*}$ given in~\eqref{closed form optimal strategy},
we have~$\T^{\updownarrow*}_{T+}=x$ as desired.
This follows from the formula for~$\de^\updownarrow$.
Let us further observe that~$\T^{\updownarrow*}$ corresponds to the deviation process
\begin{equation}
\label{eq:zspm dev}
D^{\updownarrow*}_0=\de,\quad
D^{\updownarrow*}_t=\de^\updownarrow f_t\text{ on }(0,T],\quad
D^{\updownarrow*}_{T+}=\de^\updownarrow,
\end{equation}
which immediately follows from~\eqref{Theta via D}
(direct computation using~\eqref{D zero spread} is somewhat longer).
A straightforward calculation gives
\begin{equation}
\label{eq:zspm opt val 1}
J^\updownarrow(\de,\T^{\updownarrow*})
=(\de^\updownarrow)^2 \left(\int_0^T f_t \frac{f'_t+\rho_t f_t}{K_t} dt+\frac{f^2_0}{2 K_0}+\frac{1-f_T^2}{2 K_T} \right)-\frac{\de^2}{2 K_0}.
\end{equation}
Using integration by parts we get
$$
\int_0^T\frac{f_tf'_t}{K_t}\,dt=\frac12\left[\frac{f_T^2}{K_T}-\frac{f_0^2}{K_0}
+\int_0^T \frac{f_t^2}{K_t^2}K'_t\,dt\right].
$$
Substituting this into~\eqref{eq:zspm opt val 1}
we get that $J^\updownarrow(\de,\T^{\updownarrow*})$
equals the right-hand side of~\eqref{eq:zspm opt value}.

It remains to prove optimality of~$\T^{\updownarrow*}$.
Due to Lemma~\ref{cts strategies cor} it is enough to prove that $\T^{\updownarrow*}$
is optimal in the class~$\cA^{\updownarrow c}(x)$, which we do below.
In terms of~$D^{\updownarrow *}$, the corresponding trading costs are
	\begin{eqnarray*}
		J^{\updownarrow}(\de,\T^{\updownarrow *})&=& \int_{(0,T)} D^{\updownarrow *}_t d \T^{\updownarrow *}_t +\left(\de+\frac{K_0}{2} \De \T^{\updownarrow *}_0 
		\right) \De \T^{\updownarrow *}_0+\left(D^{\updownarrow *}_T+\frac{K_T}{2} \De \T^{\updownarrow*}_T \right) \De \T^{\updownarrow *}_T \\
	  &=& \int_{(0,T)} \frac{D^{\updownarrow *}_t}{K_t} dD^{\updownarrow *}_t +\int_{(0,T)} \frac{\rho_t (D^{\updownarrow *}_t)^2}{K_t} dt +\frac{(D^{\updownarrow 
	  *}_{0+})^2-\de^2}{2 K_0}+\frac{(D^{\updownarrow *}_{T+})^2-(D^{\updownarrow *}_T)^2}{2 K_T}.
	\end{eqnarray*}
	Let us now look at alternative strategies~$\hat{\T} \in \cA^{\updownarrow c}_0(x)$ with 
	corresponding~$\hat{D}=D^{\updownarrow *}+h$ and show that these alternative strategies cause higher trading costs than~$\T^{\updownarrow *}$. That is in the 
	following, we work with functions~$h\colon[0,T+] \rightarrow \R$, which are of bounded variation and continuous on~$(0,T)$ with $h_0=0$, $h_T=\lim_{t\nearrow T}h_t$ and a finite limit $h_{0+}$
(so that there are possibly jumps~$(h_{0+}-h_0),(h_{T+}-h_T) \in \R$). Using
	$$ \De \hat{\T}_0 = \De \T^{\updownarrow *}_0 + \frac{h_{0+}}{K_0}, \hspace{.5 cm}
		 d \hat{\T}_t= d\T^{\updownarrow *}_t+\frac{d h_t+\rho_t h_t dt}{K_t}, \hspace{.5 cm}
	   \De \hat{\T}_{T} = \De \T^{\updownarrow *}_{T} + \frac{h_{T+}-h_T }{K_T},$$
	a straightforward calculation yields
	\begin{eqnarray*}
		&&J^{\updownarrow}(\de,\hat{\T}) = \int_{(0,T)} \hat{D}_t d \hat{\T}_t +\left(\de+\frac{K_0}{2} \De \hat{\T}_0 \right) \De \hat{\T}_0  
		  +\left(\hat{D}_T+\frac{K_T}{2} \De \hat{\T}_T \right) \De \hat{\T}_T \\
	                 	&& \hspace*{1.65cm}= J^{\updownarrow}(\de,\T^{\updownarrow *})+ \De J_1+ \De J_2,\\
	  && \hspace*{1cm}\\
		&& \De J_1 :=	\int_{(0,T)} \frac{2 \rho_t D^{\updownarrow *}_t h_t}{K_t} dt+ \int_{(0,T)} \frac{h_t}{K_t} dD^{\updownarrow *}_t+ \int_{(0,T)} 
		\frac{D^{\updownarrow *}_t}{K_t} dh_t 
		+\frac{D^{\updownarrow *}_{0+}h_{0+}}{K_0}+\frac{D^{\updownarrow *}_{T+}h_{T+}-D^{\updownarrow *}_T h_T}{K_T},\\
		&& \De J_2 := \int_{(0,T)} \frac{\rho_t h_t^2}{K_t} dt + \int_{(0,T)} \frac{h_t}{K_t}dh_t +\frac{h^2_{0+}}{2K_0}+\frac{h^2_{T+}-h^2_T}{2 K_T}.
	\end{eqnarray*}
	Notice that we collect all terms containing~$D^{\updownarrow *}$ in~$\De J_1$. We are now going to finish the proof by showing that~$\De J_1=0$ and~$\De J_2 > 0$ 
	if~$h$ does not vanish.
	
	Let us first rewrite~$\De J_1$ exploiting the fact that~$D^{\updownarrow *}_t=\de^\updownarrow f_t$, use integration by parts, the definition of~$f$ and 
	again integration by parts to get
	\begin{eqnarray*}
		&&\De J_1 \\
		&=& \de^\updownarrow \left \{ \int_{(0,T)} \frac{2 \rho_t f_t h_t}{K_t} dt+ \int_{(0,T)} \frac{h_t}{K_t} df_t+ \int_{(0,T)} \frac{f_t}{K_t} dh_t  
		  +\frac{f_{0}h_{0+}}{K_0}+\frac{h_{T+}-f_T h_T}{K_T} \right\} \\
		        &=& \de^\updownarrow \left \{ \int_{(0,T)} \frac{2 \rho_t K_t+K'_t}{K^2_t}f_t h_t dt +\frac{h_{T+}}{K_T} \right\} \\
		        &=& \de^\updownarrow \left \{ \int_{(0,T)} \frac{\rho_t h_t}{K_t} dt +\frac{h_{T+}}{K_T} +\int_{(0,T)} \frac{K'_t}{K_t} h_t dt \right\} \\
		        &=& \de^\updownarrow \left\{ \int_{(0,T)} \frac{\rho_t h_t}{K_t}dt + \int_{(0,T)} 
		        \frac{1}{K_t}dh_t+\frac{h_{0+}}{K_0}+\frac{h_{T+}-h_T}{K_T}\right\}.
	\end{eqnarray*}	
Clearly, $\De J_1=0$ whenever $\de^\updownarrow=0$.
If $\de^\updownarrow\ne0$, we have
	\begin{eqnarray*}
		x &=& \int_{(0,T)} d \hat{\T}_t+\De \hat{\T}_0+\De \hat{\T}_T\\
		  &=& \left(\int_{(0,T)} d \T^{\updownarrow *}_t + \De \T^{\updownarrow *}_0+\De \T^{\updownarrow *}_T\right) 
		  + \left(\int_{(0,T)} \frac{\rho_t h_t}{K_t}dt + \int_{(0,T)} \frac{1}{K_t}dh_t +\frac{h_{0+}}{K_0}+\frac{h_{T+}-h_T}{K_T}\right)\\
		  &=&x + \frac{\De J_1 }{\de^\updownarrow}.
	\end{eqnarray*}
	Therefore~$\De J_1 =0$.
	Hence,~$J^{\updownarrow}(\de,\hat{\T})-J^{\updownarrow}(\de,\T^{\updownarrow*})= \De J_2$. Applying integration by parts to the~$d h_t$ integral yields
	$$\De J_2 = \int_{(0,T)} \frac{h_t^2}{2 K_t}\left(2\rho_t+ \frac{K'_t}{K_t} \right)dt + \frac{h_{T+}^2}{2K_T}.$$	
	Due to the assumption $K'_t+2\rho_tK_t>0$ on~$[0,T]$, we get that $\De J_2$ is positive as desired.
\end{proof}

\section{Examples}\label{Subsec: Euler Lagrange}

Let us now turn to explicit examples of dynamics of the price impact parameter~$K$ and the resilience~$\rho$.
We can use the formulas derived in the previous section to calculate optimal trading strategies
in problem~\eqref{eq:zspm value fun} in the zero spread model.
We also want to investigate optimal strategies
in problem~\eqref{U} in the dynamic spread model introduced in Section~\ref{SecModel}.
In~\eqref{U} we considered a general initial time $t\in[0,T]$.
Without loss of generality below we will consider initial time $0$ for both models,
e.g.~we will mean the function $U(0,\cdot,\cdot)$ when speaking about the value function in the dynamic spread model.
Further, in the dynamic spread model we had a nonnegative initial value~$\de$
for the deviation of the best ask price from its unaffected level
and considered strategies with the overall goal to buy a nonnegative number of shares~$x$.
That is, we will consider $\de,x\in[0,\infty)$ in this section when speaking about either model.
It is clear that strategy~\eqref{closed form optimal strategy} is optimal also in the dynamic spread model
whenever it does not contain selling.
Thus, Theorem~\ref{Thm zero spread opt strat}, applied with $\de,x\in[0,\infty)$, provides us with formulas
for the value function and optimal strategy also in the dynamic spread model
whenever there is no transaction-triggered price manipulation in the zero spread model
(see~Corollary~\ref{cor:zspm ttpm}) and $\de$ is sufficiently close to~$0$
(so that $\De\T^{\updownarrow*}_0$ given by the first formula in~\eqref{closed form optimal strategy}
is still nonnegative). Furthermore, in this case we get an explicit formula
for the barrier function of Definition~\ref{WR-BR structure definition}.

\begin{proposition}\label{Euler-Lagrange barrier thm}{\rm (Closed form optimal barrier in the dynamic spread model).\\}
Assume the dynamic spread model of Section~\ref{SecModel}
and that~$K\colon[0,T]\to(0,\infty)$ is twice continuously differentiable
and~$\rho\colon[0,T]\to(0,\infty)$ is continuously differentiable.
Let
\begin{equation}
\label{eq:ex no ttpm}
K'_t+2\rho_tK_t>0\text{ on }[0,T],\quad
f_0\ge0\quad\text{and}\quad
f'_t+\rho_tf_t\ge0\text{ on }[0,T],
\end{equation}
where $f$ is defined in~\eqref{eq:zspm f eq}.
Then the barrier function of Definition~\ref{WR-BR structure definition} is explicitly given by
\begin{equation}
\label{eq:ex barrier}
c(t)=\frac{1}{f_t}\left(\int_t^T \frac{f'_s+\rho_s f_s}{K_s}ds+\frac{1-f_T}{K_T}\right),\quad t\in[0,T),
\quad c(T)=0.
\end{equation}
Furthermore, for any $x\in[0,\infty)$ and $\de\in\left[0,\frac x{c(0)}\right]$,
there is a unique optimal strategy in the problem~$U(0,\de,x)$ (see~\eqref{U})
and it is given by formula~\eqref{closed form optimal strategy} in Theorem~\ref{Thm zero spread opt strat},
and the value function $U(0,\de,x)$ equals the right-hand side of~\eqref{eq:zspm opt value}.
\end{proposition}

\begin{remark}[Comments to~\eqref{eq:ex barrier}]
\label{rem:ex barrier}\mbox{}

\vspace{-4mm}
\begin{enumerate}[i)]
\item First let us note that~\eqref{eq:ex no ttpm} implies $f_t\ge0$ on~$[0,T]$
(see Lemma~\ref{lem:ex no ttpm equiv} below). Hence the right-hand side of~\eqref{eq:ex barrier}
has the form $a/b$ with $a>0$ (note that $f_T<1$) and $b\ge0$,
i.e.~$c(t)\in(0,\infty]$ for~$t\in[0,T)$.
The case $c(t)=\infty$ can occur (see~e.g. Example~\ref{example exponential} with~$\nu=-1$).
\item Let us further observe that
$$
\lim_{t\nearrow T}c(t)=\frac{1-f_T}{f_TK_T}\in(0,\infty],
$$
i.e. the barrier always jumps at~$T$.
\end{enumerate}
\end{remark}

\begin{proof}[Proof of Proposition~\ref{Euler-Lagrange barrier thm}]
Let us first notice that~$c$ from~\eqref{eq:zspm c greater 0} is strictly positive by Lemma~\ref{lemma c>0},
so that Theorem~\ref{Thm zero spread opt strat} applies.
Further, it follows from~\eqref{eq:ex no ttpm} that in the zero spread model
with such functions~$K$ and $\rho$ there is no transaction-triggered price manipulation.
Hence, for any~$x>0$, the optimal strategy $\T^{\updownarrow*}$ from~\eqref{closed form optimal strategy} with $\de=0$
in the problem~$U^\updownarrow(0,x)$ will also be optimal in the problem~$U(0,0,x)$.
Let us recall that the value~$c(0)$ of the barrier is the ratio $\frac{x-\De\T^{\updownarrow*}_0}{D^{\updownarrow*}_{0+}}$
for the optimal strategy $\T^{\updownarrow*}$ in the problem~$U(0,0,x)$
and the corresponding~$D^{\updownarrow*}$ (with~$D^{\updownarrow*}_0=0$).
Thus, we get
$$c(0)=\frac{x-\De \T^{\updownarrow *}_0}{K_0 \De \T^{\updownarrow *}_0}
=\frac{1}{f_0}\left(\int_0^T \frac{f'_s+\rho_s f_s}{K_s}ds+\frac{1-f_T}{K_T}\right).$$
A similar reasoning applies to an arbitrary~$t\in [0,T)$.
Recall that we always have $c(T)=0$.
Finally, for~$\de>0$, under condition~\eqref{eq:ex no ttpm},
formula~\eqref{closed form optimal strategy} for the zero spread model
will give the optimal strategy in the problem~$U(0,\de,x)$
(i.e.~for the dynamic spread model) if and only if $\De\T^{\updownarrow*}_0\ge0$.
Solving this inequality with respect to~$\de$ we get $\de\le\frac x{c(0)}$
(note that $\de^\updownarrow$ from~\eqref{closed form optimal strategy} also depends on~$\de$).
\end{proof}

Condition~\eqref{eq:ex no ttpm} ensures the applicability of Theorem~\ref{Thm zero spread opt strat}
and additionally excludes transaction-triggered price manipulation in the zero spread model
(see~Corollary~\ref{cor:zspm ttpm}).
The following result provides an equivalent form for this condition,
which we will use below when studying specific examples.

\begin{lemma}[An equivalent form for condition~\eqref{eq:ex no ttpm}]
\label{lem:ex no ttpm equiv}\mbox{}\\
Assume that $K\colon[0,T]\to(0,\infty)$ is twice continuously differentiable
and $\rho\colon[0,T]\to(0,\infty)$ is continuously differentiable.
Then condition~\eqref{eq:ex no ttpm} is equivalent to
\begin{equation}
\label{eq:ex no ttpm equiv}
K'_t+\rho_tK_t\ge0\text{ on }[0,T]
\quad\text{and}\quad
f'_t+\rho_tf_t\ge0\text{ on }[0,T].
\end{equation}
\end{lemma}

\begin{proof}
Clearly, \eqref{eq:ex no ttpm equiv} implies~\eqref{eq:ex no ttpm}.
Let us prove the converse.
Suppose \eqref{eq:ex no ttpm}~is satisfied and ${K'_s+\rho_sK_s<0}$ for some $s\in[0,T]$.
Then there exists $[u,v]\subset[0,T]$ such that $u<v$, $f_u=0$ and $f_t<0$ on~$(u,v]$.
By the mean value theorem, there exists $w\in(u,v)$ such that $f'_w=(f_v-f_u)/(v-u)$.
Thus, we get $f'_w<0$ and $f_w<0$, which contradicts the condition $f'_t+\rho_tf_t\ge0$ on~$[0,T]$.
\end{proof}

When we have transaction-triggered price manipulation in the zero spread model,
optimal strategies in the dynamic spread model are different from the ones
given in Theorem~\ref{Thm zero spread opt strat}.
The following proposition deals with the case of $K'_t+\rho_t K_t<0$ for some~$t$
(cf.~with~\eqref{eq:ex no ttpm equiv}).

\begin{proposition}\label{Euler lemma 1}{\rm (Wait if decrease of~$K$ outweighs resilience).\\}
Assume the dynamic spread model of Section~\ref{SecModel}.
Let, for some~$t \in [0,T)$, $K$ be continuously differentiable at~$t$ and $\rho$ continuous at~$t$ with~$K'_t+\rho_t K_t<0$. Then $Br_t= \emptyset$, i.e., $c(t)=\infty$.
\end{proposition}

\begin{proof}
Since $K'+\rho K$ is continuous at~$t$, we have $K'_s+\rho_s K_s<0$ on an interval around~$t$.
Then there exists~$\ep>0$ such that~$K_s e^{-\int_s^{t+\ep}\rho_u du} >K_{t+\ep}$ for all $s \in [t,t+\ep)$.
By Proposition~\ref{lemma infinite barrier}, it is not optimal to trade at~$t$. 
\end{proof}

Let us finally illustrate our results by discussing several examples.
For simplicity, take constant resilience~$\rho>0$.
Then condition~\eqref{eq:ex no ttpm equiv} takes the form
\begin{equation}\label{Euler cond constant rho}
K'_t+\rho K_t\ge0\text{ on }[0,T]
\quad\text{and}\quad
K''_t+3\rho K'_t+2\rho^2 K_t\ge0\text{ on }[0,T].
\end{equation}
A sufficient condition for~\eqref{Euler cond constant rho},
which is sometimes convenient (e.g.~in Example~\ref{example exponential} below), is
$$
K'_t+\rho K_t\ge0\text{ on }[0,T]
\quad\text{and}\quad
K''_t+\rho K'_t\ge0\text{ on }[0,T].
$$
In all examples below we consider $\de=0$ and $x\in[0,\infty)$.

\begin{example}\label{example constant}{\rm (Constant price impact~$K_t \equiv \ka>0$).\\}
	Assume that the price impact~$K_t \equiv \ka>0$ is constant. 
	Clearly, condition~\eqref{Euler cond constant rho} is satisfied,
	so we can use formula~\eqref{closed form optimal strategy}
	to get the optimal strategy in both models.
	We have $f_t \equiv \frac{1}{2}$ and $\de^\updownarrow=\frac{2\ka x}{\rho T+2}$.
	The optimal strategy in both the dynamic and zero spread models is given by the formula
  $$ \De \T_0=\De \T_T=\frac{x}{\rho T+2}, \quad
  d \T_t =\frac{x\rho}{\rho T+2}\,dt,$$
	which recovers the results from~\citeasnoun{OW}. 
	The large investor trades with constant speed on~$(0,T)$ and consumes all fresh limit sell orders entering the book 
	due to resilience in such a way that the corresponding deviation process~$D_t$ is constant on~$(0,T]$
	(see~\eqref{eq:zspm dev} and note that $f_t$ is constant). 
	The barrier is linearly decreasing in time (see~\eqref{eq:ex barrier}):
	$$c(t)=\frac{1+\rho (T-t)}{\ka},\quad t\in[0,T),\quad c(T)=0.$$
	Let us finally note that the optimal strategy does not depend on~$\ka$,
	while the barrier depends on~$\ka$.
	See Figure~\ref{OW figure} for an illustration.
\begin{figure}[htbp]
 \centering
 \includegraphics[width=0.49\linewidth]{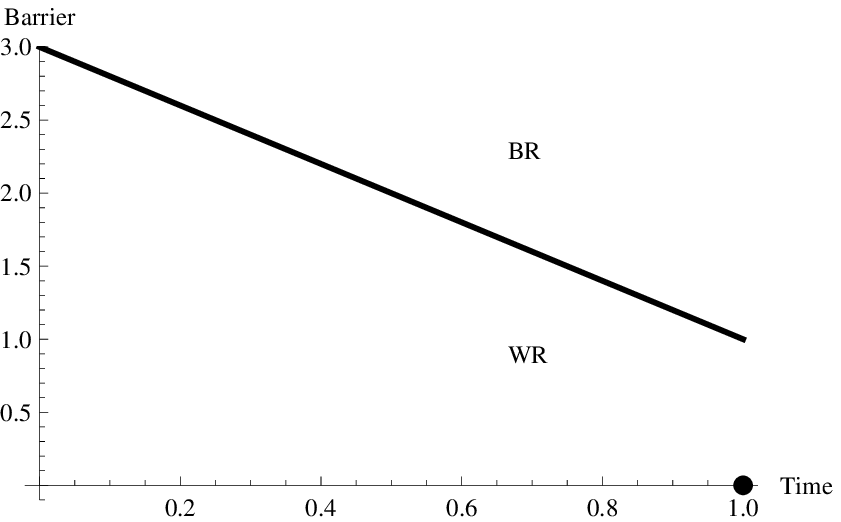}
 \includegraphics[width=0.49\linewidth]{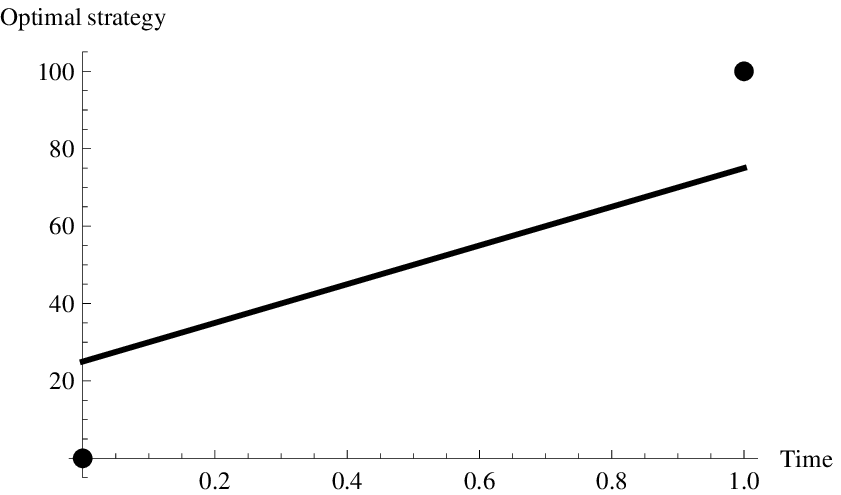}
 \caption{Constant price impact ($T=1, \rho=2,\ka=1,x=100,\de=0$).
 }
 \label{OW figure}
\end{figure}
\end{example}

\begin{example}\label{example exponential}{\rm (Exponential price impact
$K_t=\ka e^{\nu \rho t}$, $\ka>0$, $\nu\in\R\setminus\{0\}$).\\}
Assume that the price impact~$K_t=\ka e^{\nu \rho t}$ is growing or falling exponentially
with $\nu \in \R\setminus\{0\}$ being the slope of the exponential price impact relative to the resilience.
The case $\nu=0$ was studied in the previous example.
We exclude this case here because some expressions below will take the form $0/0$ when $\nu=0$
(however, the limits of these expressions as $\nu\to0$ will recover the corresponding formulas from the previous example).
Condition~\eqref{Euler cond constant rho} is satisfied if and only if~$\nu \geq -1$. 
We first consider the case $\nu\ge-1$.
We have
$$
f_t\equiv\frac{\nu+1}{\nu+2}\quad\text{and}\quad
\de^\updownarrow=\frac{x\ka\nu(\nu+2)}{(\nu+1)^2-e^{-\nu\rho T}}.
$$
In particular, like in the previous example, the large investor trades in such a way
that the deviation process $D_t$ is constant on~$(0,T]$.
The optimal strategy in both the dynamic and zero spread models is given by the formula
$$
\De \T_0=\frac{ x \nu(\nu+1)}{(\nu+1)^2-e^{-\nu\rho T}},\quad
d\T_t=\frac{ x \nu(\nu+1)}{(\nu+1)^2-e^{-\nu\rho T}}\,\rho e^{-\nu \rho t}\,dt,\quad
\De \T_T=\frac{ x \nu }{(\nu+1)^2-e^{-\nu\rho T}}e^{-\nu \rho T}.
$$
We see that, for~$\nu=-1$, it is optimal to buy the entire order at~$T$.
Vice versa, the initial trade $\De\T_0$ approaches~$x$ as~$\nu\nearrow\infty$.
The barrier is given by the formula
$$
c(t)=\frac{(\nu+1)e^{-\nu\rho t}-e^{-\nu\rho T}}{\ka\nu(\nu+1)},\quad t\in[0,T),\quad c(T)=0
$$
(in particular, $c(t)=\infty$ for~$t\in[0,T)$ if~$\nu=-1$ and the barrier is finite everywhere if~$\nu>-1$).
For each~$\nu>-1$, the barrier is decreasing in~$t$, i.e.~buying becomes more aggressive as the investor runs out of time.
Furthermore, one can check that, for each $t\in[0,T)$, the barrier is decreasing in~$\nu$.
That is, the greater is~$\nu$, the larger is the buy region since it is less attractive to wait.
Like in the previous example, the optimal strategy does not depend on~$\ka$, while the barrier depends on~$\ka$.

Let us now consider the case $\nu<-1$.
In the zero spread model, transaction-triggered price manipulation occurs for~$\nu\in(-2,-1)$
(one checks that the assumptions of Theorem~\ref{Thm zero spread opt strat} are satisfied)
and classical price manipulation occurs for~$\nu<-2$ (see~Proposition~\ref{Thm zero spread pm}).
In the dynamic spread model, for $\nu<-1$,
it is optimal to trade the entire order at~$T$
because~$K_t e^{-\rho(T-t)}>K_T$ for all~$t \in [0,T)$
(see~Proposition~\ref{lemma infinite barrier}).
Thus, in the case~$\nu<-1$,
we have $c(t)=\infty$ for~$t\in[0,T)$.

See Figure~\ref{exponential figure} for an illustration.
	\begin{figure}[htbp]
	 \centering
	 \includegraphics[width=0.4\linewidth]{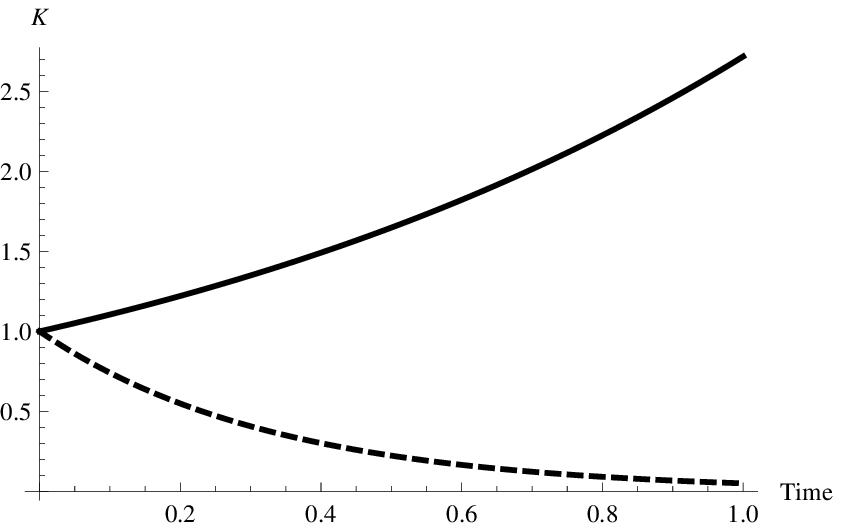}
	 \includegraphics[width=0.4\linewidth]{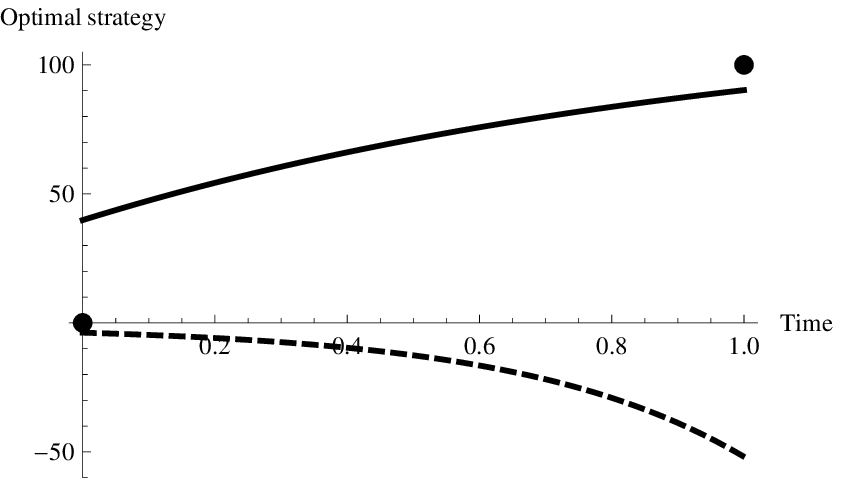}
   \caption{Exponential price impact ($T=1, \rho=2,\ka=1,x=100,\de=0,\nu=0.5$ and~$-1.5$ (dashed)).
	 }
	 \label{exponential figure}
	\end{figure}
\end{example}

\begin{example}\label{example straight line}{\rm (Straight-line price impact $K_t=\ka +m t$, $\ka>0$, $m>-\frac\ka T$).\\}
Assume that the price impact $K_t=\ka +m t$ changes linearly over time.
The condition $m>-\frac{\ka}{T}$ ensures that $K$ is everywhere strictly positive.
Condition~\eqref{Euler cond constant rho} is satisfied if and only if $m\geq-\frac{2\rho\ka}{3+2\rho T}$.
Note that $-\frac{2\rho\ka}{3+2\rho T}>-\frac\ka T$.
Let us first assume that $m\geq-\frac{2\rho\ka}{3+2\rho T}$.
In this case, the optimal strategy in both the dynamic and zero spread models is given by the formulas
\begin{eqnarray*}
\De \T_0 &=& \frac{2m(m+\ka \rho)x}{(m+2 \ka \rho) \tilde{m}},\hspace{1.5cm} d\T_t=\frac{2m \ka \rho^2\left(2 \ka \rho+m(3+2 \rho t)\right)x}{\left(m+2 \ka \rho+2m \rho t\right)^2 \tilde{m}}dt,\\
\De \T_T &=& \frac{2m \ka \rho x}{(m+2 \ka \rho+2m \rho T) \tilde{m}} \hspace{.5cm} \text{with} \hspace{.2cm} \tilde{m}:=2m+\ka \rho \log\left(\frac{m+2 \ka \rho+2 m \rho T}{m+2 \ka \rho}\right).
\end{eqnarray*}
The barrier is given by the formula
$$c(t)=\rho\frac{ 2 m - (m + 2 \ka \rho + 2 m \rho t) \log\left( \frac{m + 2 \ka \rho + 2 m \rho t}{m + 2 \ka \rho + 2 m \rho T}\right)}{2 m (m + \ka \rho + m \rho t)}.$$
In the zero spread model, transaction-triggered price manipulation occurs
for $m\in(-\frac{2\rho\ka}{1+2\rho T},-\frac{2\rho\ka}{3+2\rho T})$ (see Theorem~\ref{Thm zero spread opt strat})
and classical price manipulation occurs for $ m\in(-\frac\ka T,-\frac{2\rho\ka}{1+2\rho T})$ (see Proposition~\ref{Thm zero spread pm}).
In the dynamic spread model, we can check by Proposition~\ref{lemma infinite barrier}
that it is optimal to trade the entire order at~$T$ for
$$
m\in\left(-\frac\ka T,-\frac\ka T\left(1-e^{-\rho T}\right)\right)
$$
(see~Lemma~\ref{lem:app_b2}).
We observe that $-\frac\ka T\left(1-e^{-\rho T}\right)<-\frac{2\rho\ka}{3+2\rho T}$ (see~Lemma~\ref{lem:app_b1}).
Let us finally note that the presented methods do not allow us to calculate the optimal strategy in closed form
in the dynamic spread model for
$m\in\left[-\frac\ka T\left(1-e^{-\rho T}\right),-\frac{2\rho\ka}{3+2\rho T}\right)$,
but we can approximate it numerically in discrete time
(see~e.g. the case with~$K_t=1-0.6t$, $\rho=2$, $T=1$ in Figure~\ref{Discrete figure}).

See Figure~\ref{straight line figure} for an illustration.
	\begin{figure}[htbp]
	 \centering
	 \includegraphics[width=0.4\linewidth]{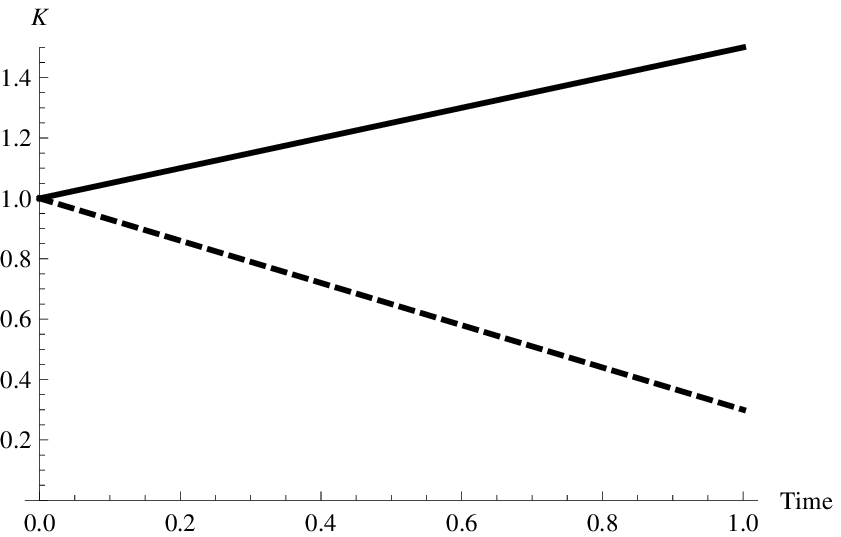}
	 \includegraphics[width=0.4\linewidth]{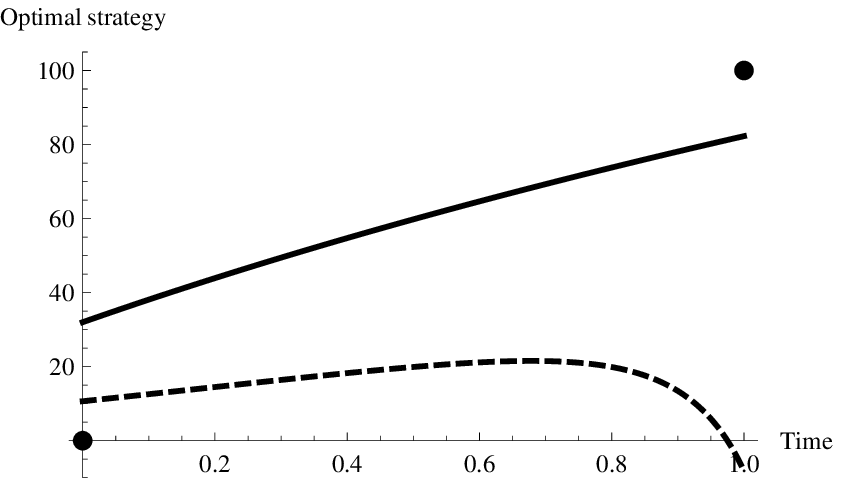}
	 \caption{Straight-line price impact ($T=1,\rho=2,\ka=1,x=100,\de=0,m=0.5$ and~$-0.7$ (dashed)).
	 }
	 \label{straight line figure}
	\end{figure}
\end{example}

\section{Conclusion}\label{sec:conclusion}
Time-varying liquidity is a fundamental property of financial markets. Its implications for optimal liquidation in limit order book markets is the focus of this paper. We find that a model with a dynamic, trading influenced spread is very robust and free of two types of price manipulation. We prove that value functions and optimal liquidation strategies in this model are of wait-region/buy-region type, which is often encountered in problems of singular control.
In the literature on optimal trade execution in limit order books, the spread is often assumed to be zero. Under this assumption we show that time-varying liquidity can lead to classical as well as transaction-triggered price manipulation. For both dynamic and zero spread assumptions we derive closed form solutions for optimal strategies and provide several examples.

\appendix

\section{Integration by parts for c\`agl\`ad processes}\label{AppendixStochAna}
In various proofs in this paper we need to apply stochastic analysis (e.g. integration by parts or Ito's formula) to c\`agl\`ad processes of finite variation and/or standard semimartingales. As noted in Section~\ref{SecModel}, this is always done as follows: if~$U$ is a c\`agl\`ad process of finite variation, we first consider the process~$U^+$ defined by~$U^+_t:=U_{t+}$ and then apply standard formulas from stochastic analysis to it. As an example of such a procedure we provide the following lemma, which is often applied in the proofs in this paper. 
\begin{lemma}\label{int by parts lemma}{\rm (Integration by parts).\\}
	Let~$U=(U_t)_{t \in [0,T+]}$ and~$V=(V_t)_{t \in [0,T+]}$ be c\`agl\`ad processes of finite variation and~$Z$ a semimartingale (in particular  
	c\`adl\`ag), which may have a jump at~$0$. For~$t \in [0,T]$, we have
	\begin{eqnarray}
		\label{int by parts 1}
		U_{t+} Z_t &=& U_0 Z_{0-}+\int_{[0,t]} U_s dZ_s + \int_{[0,t]}Z_s dU_s,\\
		\label{int by parts 2}
		U_{t+} V_{t+} &=& U_0 V_0+\int_{[0,t]} U_s dV_s + \int_{[0,t]}V_{s+} dU_s.
	\end{eqnarray}
\end{lemma}

\begin{proof}
Let~$X$ and~$Y$ be c\`adl\`ag processes (possibly having a jump at~$0$) with~$X$ being a semimartingale and~$Y$ a finite variation process. By Proposition I.4.49 a) in \citeasnoun{JacodShiryaev}, which is a variant of integration by parts for the case where one of the semimartingales is of finite variation,
\begin{equation}\label{int by parts JS}
	X_t Y_t = X_{0-} Y_{0-}+ \int_{[0,t]} Y_{s-}dX_s+ \int_{[0,t]}X_sdY_s, \hspace{.2cm} t \in [0,T].
\end{equation}
Equation~(\ref{int by parts 1}) is a particular case of~(\ref{int by parts JS}) applied to~$X:=Z, \, Y:=U^+$ and equation~(\ref{int by parts 2}) is a particular case of~(\ref{int by parts JS}) applied to~$X:=V^+, \, Y:=U^+$, where~$U^+_t:=U_{t+}$ and~$V^+_t:=V_{t+}$.
\end{proof}

\section{Technical lemmas used in Example~\protect\ref{example straight line}}
\label{app:b}
Below we use the notation of Example~\ref{example straight line}.

\begin{lemma}
\label{lem:app_b2}
For $m\in\left(-\frac\ka T,-\frac\ka T\left(1-e^{-\rho T}\right)\right)$ we have
\begin{equation}\label{straight line inequality}
	(\ka+mt)e^{-\rho(T-t)}>\ka+mT,\quad t\in[0,T),
\end{equation}
i.e. Proposition~\ref{lemma infinite barrier} applies.
\end{lemma}

\begin{proof}
Inequality~(\ref{straight line inequality}) is equivalent to
$$m<-\frac{\ka \left(1-e^{-\rho(T-t)}\right)}{T-t e^{-\rho(T-t)}}.$$
The assertion now follows from our assumption on~$m$. To see this, we need to show that
$$-\frac\ka T\left(1-e^{-\rho T}\right) \leq -\frac{\ka \left(1-e^{-\rho(T-t)} \right)}{T-t e^{-\rho(T-t)}},$$
which in turn is equivalent to
$$g(t):=\frac{1-e^{-\rho(T-t)}}{T-t e^{-\rho(T-t)}} \leq \frac{1-e^{-\rho T}}{T}=g(0).$$
This is a true statement since~$1-x \leq e^{-x}$ for all~$x \in \R$ and therefore
$$g'(t)=\frac{1-\rho (T-t)-e^{-\rho(T-t)}}{e^{\rho(T-t)}(T-t e^{-\rho(T-t)})^2}\leq 0.$$
\end{proof}

\begin{lemma}
\label{lem:app_b1}
We have $-\frac\ka T\left(1-e^{-\rho T}\right)<-\frac{2\rho\ka}{3+2\rho T}$.
\end{lemma}

\begin{proof}
The statement reduces to proving that $\frac{2\rho T}{3+2\rho T}<1-e^{-\rho T}$.
Setting $x:=\rho T>0$ we see that it is enough to establish that $e^{-x}<\frac3{3+2x}$,
which is true as, clearly, $e^x>1+\frac23 x$.
\end{proof}

\addcontentsline{toc}{section}{Bibliography}
\bibliographystyle{jf}
\bibliography{References}
\end{document}